\documentclass[11pt]{article}

\newfont{\tenmsb}{msbm10 scaled\magstep1}

\let\ssection=\section
\renewcommand{\section}{\setcounter{equation}{0}\ssection}
\usepackage{graphics,amssymb,amsmath,amsthm,amscd,array}
\usepackage{graphicx}
\usepackage{makeidx}

\font\small=cmr7

\def\parag{\hfil\break} %%%%% paragraph
\def\kikezd{\parag\underbar}
\def\IR{{\bf R}} %%%%% Reals
\def\L{{\cal L}}

\def\smallover#1/#2{\hbox{$\textstyle\frac{#1}{#2}$}} %
\def\smallcirc{{\raise 0.5pt \hbox{$\scriptstyle\circ$}}}
\def\2{{\smallover1/2}}
\def\={{\!=\!}}
\def\O{{\rm O}}
\def\o{{\rm o}}

\def\p{{\partial }}
\def\dAlembert{\vcenter {
    \hbox {\vrule height8pt width0.4pt depth0.0pt
           \vrule height8pt width7.2pt depth-7.6pt
           \vrule height8pt width0.4pt depth0.0pt
           \kern-8pt
           \vrule height0.4pt width8pt depth0.0pt
          }}}%--- Francisco's box 

\def\and{\qquad\hbox{and}\qquad}
\def\where{\qquad\hbox{where}\qquad}

\def\rot{{\rm curl\ }}
\def\grad{{\rm grad\ }}

\def\vnabla{{\vec{\nabla}}}
\def\vJ{{\vec{J}}}
\def\vD{{\vec{D}}}
\def\vA{{\vec{A}}}
\def\vD{{\vec{D}}}
\def\vB{{\vec{B}}}
\def\vE{{\vec{E}}}
\def\vx{{\vec{x}}}
\def\vp{{\vec{p}}}
\def\va{{\vec{a}}}
\def\I{{\rm I}}
\def\cH{{\cal H}}
\def\cP{{\cal P}}
\def\cJ{{\cal J}}
\def\cG{{\cal G}}
\def\cD{{\cal D}}
\def\cK{{\cal K}}

\def\cN{{\cal N}}

\def\D{
{D\mkern-2mu\llap{{\raise+0.5pt\hbox{\big/}}}\mkern+2mu}
}  %Diracop
\begin{document}
%%%%%%%%%%%%%%%%%%

\title{Lectures on (abelian) Chern-Simons vortices
}

\author{
P.~A.~Horv\'athy
\\
Laboratoire de Math\'ematiques et de Physique Th\'eorique\\
Universit\'e de Tours\\
Parc de Grandmont\\
F-37 200 TOURS (France)
\\ 
}

\maketitle

\begin{abstract}
Various aspects including the construction and
the symmetries of Abelian Chern-Simons vortices
are reviewed. Extended version of the Lectures
delivered at NIKHEF (Amsterdam), July 2006. 
\end{abstract}

\bigskip
\texttt{ArXiv: 0704.3220}
    
\setlength{\baselineskip}{15pt}

%%%%%%%%%%%%%%%%%%%%%%%%%%%%%%%%%%%%%%%%%%%
%%%%%%%%%%%%%%%% the text %%%%%%%%%%%%%%%%%
%%%%%%%%%%%%%%%%%%%%%%%%%%%%%%%%%%%%%%%%%%%

\newpage\null\newpage
\tableofcontents
\newpage\null\newpage

%%%%%%%%%%%%%%%%%%%%%%%%%%%%%%%%%%%%%%%%%%%%%%%%
\section{Introduction~: the Chern-Simons form.
}
%%%%%%%%%%%%%%%%%%%%%%%%%%%%%%%%%%%%%%%%%%%%%%%%%

The interaction between physics and mathematics can go in
both ways.  For example, Maxwell's theory, introduced to describe  
 electromagnetism, has later been applied also to
mathematics, namely to potential theory.
This happened again when Maxwell's electromagnetism was generalized to describe non-Abelian interactions -- and Yang-Mills theory
became, later, an essential tool in differential
geometry for studying the characteristic
(Pontryagin) classes over even-dimensional manifolds.

With Chern-Simons theory history went the opposite way~:
in the early 1970,
S. S. Chern and  Simons \cite{Chern} introduced their
secondary characteristic classes to study 
bundles over odd-dimensional manifolds; this geometrical
tool found subsequent application in low-dimensional
physics. In $3$ space-time dimensions (the only case we study in 
this Review), the (Abelian) Chern-Simons three-form is\footnote{
Three-dimensional space-time indices are denoted by
$\alpha, \beta,\dots$
The non-Abelian generalization (not considered here) 
of the Chern-Simons form  is
$$
\frac{1}{4}%\epsilon^{\alpha\beta\gamma}
{\rm Tr}\left(
A_\alpha F_{\beta\gamma}
-\frac{2}{3}A_\alpha A_\beta A_\gamma\right)
dx^\alpha\wedge dx^\beta\wedge dx^\gamma.
$$}
\begin{equation}
\hbox{(CS form)}=
\frac{1}{4}%\epsilon^{\alpha\beta\gamma}
A_\alpha F_{\beta\gamma}\,
dx^\alpha\wedge dx^\beta\wedge dx^\gamma
\label{CSform}
\end{equation}
where $A=A_\alpha dx^\alpha$ is some vector potential.

The first applications of the Chern-Simons form to physics
 came in the early 1980, namely in
\textit{topologically massive gauge theory} \cite{CSem,DJT}. It has been
realized that (\ref{CSform}) can indeed be added to
 the usual Maxwell term in the electromagnetic action
\begin{equation}
S=S_{em}+S_{CS}=\int \left(\frac{1}{4}F_{\alpha\beta}F^{\alpha\beta}-
\frac{\kappa}{4}\epsilon^{\alpha\beta\gamma}A_\alpha F_{\beta\gamma}\right)
d^3x.
\label{CSemAction}
\end{equation}

A novel feature is that while the 3-form
(\ref{CSform}) is \textit{not} invariant under a gauge transformation $A_\alpha\to A_\alpha+\p_\alpha\lambda$,
$$ %\begin{equation}
\hbox{(CS form)}\to\hbox{(CS form)}-
\frac{\kappa}{4}
\epsilon^{\alpha\beta\gamma}(\p_\alpha\lambda)F_{\beta\gamma}\,
dx^\alpha\wedge dx^\beta\wedge dx^\gamma,
$$ %\end{equation}
the field equations associated with $S_{CS}$,
\begin{equation}
\p_\alpha F^{\alpha\gamma}+\frac{\kappa}{2}
\epsilon^{\alpha\beta\gamma}F_{\alpha\beta}=0,
\label{CSFE}
\end{equation}
\textit{are} gauge invariant. This is understood by noting that,
using the sourceless Maxwell equation
$\epsilon^{\alpha\beta\gamma}\p_\alpha F_{\beta\gamma}=0$,
the Lagrangian action (\ref{CSemAction}) is seen
to change by a mere surface term,
$$ %\begin{equation}
\Delta L_{CS}=
-
\p_\alpha\left(\frac{\kappa}{4}
\epsilon^{\alpha\beta\gamma}F_{\beta\gamma}\lambda\right)
$$ %\end{equation}
and defines, therefore, a fully satisfactory gauge theory in
3 dimensions.  Moreover,
it can be inferred from (\ref{CSFE}) that the Chern-Simons
dynamics endows the gauge field $A_\mu$ with a ``topological
mass'' \footnote{The 
Chern-Simons form behaves in a way analogous to what happens for a
Dirac monopole, for which no global vector potential 
exists, but the classical action is, nevertheless, satisfactorily defined.
In the non-Abelian context and over a compact space-time
manifold, this leads,  in a way analogous to the Dirac quantization of the monopole charge,
 to the quantization of the
Chern-Simons coefficient interpreted as the topological mass,
 \cite{DJT, JCS, HAction}.}. 

The Chern-Simons term can be used, hence, 
to supplement the usual
Maxwellian dynamics; it can even replace it altogether.
The resulting dynamics is ``poorer'', since it allows
no propagating modes.  It has, in turn, larger symmetries~:
while the Maxwell term $(1/4)F_{\alpha\beta}F^{\alpha\beta}$
requires giving a metric $g_{\alpha\beta}$, the Chern-Simons 
term is \textit{topological}~:  the integral
$$
\int \frac{1}{4}\epsilon^{\alpha\beta\gamma}A_\alpha F_{\beta\gamma}
d^3x
$$
is independent of the coordinates we choose.
Thus, while the Maxwell theory has
 historically been at the very
origin of (special) relativity, the Chern-Simons term can
allow both relativistic and non-relativistic (or even mixed)
theories.

The large invariance of the Chern-Simons term
lead, in the mid-eighties, to consider a Galilean field theory \cite{Hagen}.

The main physical application of Chern-Simons gauge theory is, however,
in condensed matter physics, namely to the 
\textit{Quantum Hall Effect} \cite{Gir,ZHK}. 
The latter,
 discovered in the early eighties \cite{QHE}, says that in a thin semiconductor
 in a perpendicular magnetic field
  the longitudinal
 resistance drops to zero
 if the magnetic field takes some specific
 values, called ``plateaus''. The current, $\vec{\jmath}$ and the electric field,
$\vec{e}$, should satisfy in turn an ``off-diagonal'' relation of the form
\begin{equation}
\vec{\jmath}=
\left(\begin{array}{ll}
0 &-\kappa\\
\kappa&0\end{array}\right)\,\vec{e}.
\label{HallLaw}
\end{equation}
where the real coefficient $\kappa$, identified as
the \textit{Hall conductivity}, is, furthermore, 
{\it quantized}.
 In the integer Quantum Hall Effect
(IQHE) $\kappa$ is an integer multiple of
some basic unit, while in the Fractional Quantum Hall Effect
(FQHE), it is a rational multiple.

The  explanation of this surprising and unexpected
quantization, provided by Laughlin's  ``microscopic'' theory,
 involves quasiparticles and quasiholes. 
 (For details the reader is referred to the literature \cite{QHE}). 
These are
composite objects that carry both  (fractional) electric charge and
a magnetic flux~: they are \textit{charged vortices}.
 
The similarities of the Fractional Quantum Hall Effect
(FQHE) with superfluidity lead condensed matter physicsist,
however,
 to ask for a phenomenological effective theory  of the FQHE \cite{Gir}.  Remember that the phenomenological description of `ordinary' superconductivity and superfluidity
is  provided by Lan\-dau-\-Ginzburg theory \cite{LiPi}~: the Cooper 
pairs formed by the electrons are represented by a scalar field,
whose charge is twice that of the electron.
The scalar fields interact through their electromagnetic fields, 
governed by the Maxwell equations. The theory admits static,
finite-energy, vortex-like solutions \cite{Abrik}. 

Landau-Ginzburg theory does not involve the time as it
``lives'' in space alone. Its relativistic extension, called 
the Abelian Higgs model \cite{NiOl},
admits again static and purely magnetic vortex-type solutions
\cite{Taubes}. Note that ordinary Landau-Ginzburg does not
admit any interesting non-relativistic
extension, owing to the intrinsically relativistic character
of the Maxwell dynamics.
 
Any ``Landau-Ginzburg'' theory  of the FQHE must reproduce Hall's law
(\ref{HallLaw}).
Now, as first pointed out in \cite{HallCS}, adding the usual
current term  $j^\alpha A_\alpha$  to the action
and suppressing the conventional Maxwell term,  
the spatial component of field equations become
precisely \textit{Hall law}, (\ref{HallLaw}). 
 This observation does not seem to have 
influenced condensed matter physicsist, though, who went in their
own way to arrive, independently, at similar conclusions.

The evolution of Chern-Simons gauge theories has been 
parallel and (almost) unrelated in 
high-energy/mathematical physics and in condensed matter physics
for at least a decade.
It is interesting to compare the early progress in  both fields~:
similar ideas arose, independently and almost simultaneously, see Table \ref{tableau}. The main difference has been
that while condensed matter physicist were more interested in
the physical derivation and in its application to the Hall effect,
high-energy/mathematical physicists explored the existence and
the construction of solutions.

\begin{table}[thp]
\begin{tabular}{|l|l|}
\hline
FIELD THEORY (hep-th)
&CONDENSED MATTER (cond-mat)\\
\hline
1981 Schonfeld; Deser, Jackiw, Templeton
&
1980-1982 v. Klitzing et al.; Tsui, Stormer, Gossard \\
%\hline
topologically massive gauge theory&Integer/Fractional Quantum
 Hall effect\\
\hline
1984-85 Hagen Galilei-invariant field&1983 Laughlin\\
 theory in 2+1d; Jackiw, Friedman et al. 
&microscopic theory of FQHE\\
relation to Hall effect
&ground-state wave functions\\
\hline
1986 Paul-Khare; De Vega-Schaposnik
&1986-87 Girvin-MacDonald
\\
vortices in Maxwell/YM + CS
&
effective `Landau-Ginzburg' theory\\
\hline
1990-91 Hong et al, Jackiw et al.
&1989 Zhang, Hansson, Kivelson
\\
relativistic/non-relativistic
&time-dependent LG theory with vortex solutions\\
topological/non-topological
&\\
self-dual vortices
&\\
\hline
1991 Ezawa et al., Jackiw-Pi
&1993 Tafelmayer\\
vortices in external field
&topological vortices in the  Zhang model\\
\hline
1997 Manton NR Maxwell-CS&\\
\hline
\end{tabular}
\caption{The Chern-Simons form in 
field theory and in condensed matter physics.}
\label{tableau}
\end{table}

The first, static, `Landau-Ginzburg' theory for the QHE has been 
put forward by Girvin \cite{Gir} on phenomenological
grounds.
An improved and extended to time-dependent theory was derived
from Laughlin's microscopic theory by Zhang, 
Hansson and Kivelson \cite{ZHK}, see Section 
\ref{LGQHE}. 

These theories involve, inevitably, the Chern-Simons form. 
In contrast to ordinary Landau-Ginzburg
theory, they can accomodate  relativistic as well as 
non-relativistic field theory is a strong argument
in its favor~: while high-energy theories are
typically relativistic, condensed matter physics is intrinsically
non-relativistic. 

Below, we review various aspects of Chern-Simons gauge theory.

In detail, we
first recall the way that lead condensed matter physicists
to  Chern-Simons theory, remarkably 
similar to that advocated by Feynman in his 
 ``Another point of view'' presented in his
1962 Lectures on Statistical Mechanics
\cite{Feynman}.

Interrupting the condensed-matter-physics approach,
the field theoretical aspects start with
Section \ref{RVort}, devoted to
 relativistic topological vortices.

Their non-relativistic limit is
physically relevant, owing to the intrinsically  non-relativistic
character of condensed matter physics. It also provides
an explicitly solvable model. For a particular choice of the potential
and for a specific value of the coupling constant, the
second-order field equations can be solved
by solving instead  first order ``self-duality'' equations.
The problem can in fact be reduced to solving the 
\textit{Liouville equation}. 
Not all solutions are physically admissible, though~: those which are
correspond to \textit{rational functions}. This provides as
with a \textit{quantization theorem of the magnetic charge},
as well as with a \textit{parameter counting}.

The symmetry problem enters the theory at (at least) two occasions. 
Firstly, do the self-dual equations come from a Bogomolny-type
decomposition of the energy~? This question becomes meaningful
if a conserved energy-momentum tensor is constructed. Such a
proceedure is canonical in a relativistic field theory, but
is rather subtle in the non-relativistic context.
 
 Another  important application %of the symmetry properties
 is to the following.
Do we have other than non-self-dual solutions at the 
specific ``self-dual'' value of the coupling constant~?
The (negative) answer is obtained  in a single
line, if the conformal symmetry of the theory is exploited
\cite{JPDT}.

Similar ideas work for vortices in a constant  background field, see Sec. \ref{background}. These models are important, since
they correspond  to those proposed 
in the Landau-Ginzburg theory of
the Fractional Quantum Hall Effect \cite{ZHK}.

Finally, we consider spinorial models. 
Again, explicit solutions are found and their symmetries are
studied using the same techniques as above.

%%%%%%%%%%%%%%%%%%%%%%%%%%%%%%%%%%%%%%%%%%%%%%
\section{Landau-Ginzburg theory for the QHE}\label{LGQHE}
%%%%%%%%%%%%%%%%%%%%%%%%%%%%%%%%%%%%%%%%%%%%%%

In Ref.~\cite{Gir}, Girvin and MacDonald call, on phenomenological grounds, for a ``Landau--Ginzburg'' 
theory for the Quantum Hall Effect.  
On phenomenological grounds, they suggest to represent the
off-diagonal long range order (ODLRO) by a scalar field $\psi(\vec{x})$ 
on the plane,
and the frustration due to deviations away from the quantized Laughlin density by
an effective gauge potential $\vec{a}(\vec{x})$.  
They propose to describe this 
static planar system by the Lagrange density
\begin{equation}
\L_{GMD}=
-\Bigl|\vec{D}\psi\Bigr|^2 
+\phi\bigl(|\psi|^2-1\bigr)
-\frac{\kappa}{2}\Bigl(\phi\; b%\vnabla\times\vec{a}
+\vec{a}\times\vnabla\phi
\Bigr),
\label{Girlag}
\end{equation}
where $b=\!\vnabla\times\vec{a}$ is the effective magnetic field,
$\vec{D}=\!\vnabla -i\,\vec{a}$ 
is the gauge-covariant derivative, and the Lagrange multiplier $\phi$ 
is a scalar potential.   
The associated equations of motion read
\begin{eqnarray}
\vec{D}^2\psi=\phi\,\psi, 
\label{matfeq}
\\[6pt]
\kappa\, b=|\psi|^2-1, 
\label{GaussFCI}
\\[6pt]
\kappa\,\vnabla\times\phi=\vec{\jmath}, 
\label{AmpHall}\vspace{-6mm}
\end{eqnarray}
where 
\begin{equation}
\vec{\jmath}=-i\big(\psi^*\vec{D}\psi-\psi(\vec{D}\psi)^*\big)
\end{equation}
is the current.

The first of these equations is a static, gauged Schr\"odinger equation
for the matter field.  

The second is the
relation proposed by Girvin and MacDonald to relate the magnetic 
field to the particle density. 

The last equation is the Amp\`ere--Hall law~: 
$\vec{e}=-i\vnabla\phi$ is an effective electric field, so that
(\ref{AmpHall}) is indeed the Hall law (\ref{HallLaw}), with
$\kappa$ identified as  the {\it Hall conductance}.

Soon after, Zhang {\it et al.\/}  \cite{ZHK}
argued that the Girvin - MacDonalds  model is merely
 a first step in the right direction and proposed 
a ``better'' Landau-Ginzburg model for the QHE,
they derive directly form the microscopic theory \cite{ZHK, ZhangRev}.
 Their starting point is the Hamiltonian
of a planar system of polarized electrons,
\begin{equation}
H_{pe}=\frac{1}{2m}\sum_a\left[\vp_a-e\vA^{ext}(\vx_a)\right]^2+
\sum_aeA_0^{ext}(\vx_a)+\sum_{a<b}V(\vx_a-\vx_b),
\label{polelHam}
\end{equation}
where $A_\alpha^{ext}$ is a vector potential
for the constant external magnetic field $B^{ext}$,
$A_i^{ext}=\2B^{ext}\epsilon_{ij}x^j$
in the symmetric gauge.
$A_0^{ext}$ is the scalar potential for the external
electric field, $E_i^{ext}=-\p_iA_0$. $V$ is the two-body
interaction potential between the electrons. The common
assumption is that $V$ is Coulombian.
The many-body wave function satisfies the
Schr\"odinger equation
\begin{equation}
H_{pe}\Psi(\vx_1,\dots,\vx_N)=E\,\Psi(\vx_1,\dots,\vx_N)
\label{peSch}
\end{equation}
and is assumed, by the Pauli principle, to be
\textit{totally antisymmetric} w.r.t. the interchange
of any two electrons.

The clue of Zhang et al. \cite{ZHK} is to map
the problem onto  a \textit{bosonic} one. Let us 
in fact consider the bosonic system with Hamiltonian
\begin{equation}
H_{bos}=\frac{1}{2m}\sum_a\left[\vp_a-e(\vA^{ext}(\vx_a)-\va(\vx_a))\right]^2+
\sum_ae\big((A_0^{ext}(\vx_a)+a_0(\vx_a)\Big)
+\sum_{a<b}V(\vx_a-\vx_b),
\label{bosHam}
\end{equation} 
where the new vector potential, $a_\alpha$, describes a
gauge interaction of specific form among the particles,
\begin{equation}
\va(\vx_a)=\frac{\Phi_0}{2\pi}\frac{\theta}{\pi} 
\sum_{b\neq a}\vnabla\gamma_{ab},
\label{statvecpot}
\end{equation}
where $\theta$ is a 
(for the moment unspecified) real parameter, 
and $\gamma_{ab}=\gamma_a-\gamma_b$ is the difference of the  
 polar angles of electrons \underbar{a} and \underbar{b}
 w.r.t. some origin and polar axis. $\Phi_0=h/ec$ is the
flux quantum. The $N$-body bosonic wave function $\phi$
is required to be \textit{symmetric} and satisfies
\begin{equation}
H_{bos}\phi(\vx_1,\dots,\vx_N)=E\,\phi(\vx_1,\dots,\vx_N).
\label{bosSch}
\end{equation}
 
Let us now  consider the singular gauge transformation
\begin{eqnarray}
\widetilde{\phi}(\vx_1,\dots,\vx_N)=U\,\phi(\vx_1,\dots,\vx_N),
\qquad
U=\exp\left[-i\sum_{a<b}\frac{\theta}{\pi}\gamma_{ab}\right].
\label{singgauge2}
\end{eqnarray}
It is easy to check that
\begin{eqnarray}
U\left[\vp_a-e(\vA^{ext}-\va)\right]U^{-1}=
\vp_a-e\vA^{ext}
\;\Longrightarrow\;
U\,H_{bos}U^{-1}=H_{pe},
\end{eqnarray}
so that $\phi$ satisfies (\ref{bosSch}) precisely
when $\widetilde{\phi}$ satisfies the polarized-electron
eigenvalue problem (\ref{peSch}) with the same eigenvalue.

To conclude our proof, let us observe that $\widetilde{\phi}$
is antisymmetric precisely when the parameter $\theta$
is an \textit{odd multiple} of $\pi$,
\begin{equation}
\theta=(2k+1)\pi.
\end{equation}

Having replaced the fermionic problem by a
bosonic one with the strange interaction
(\ref{statvecpot}), Zhang et al. proceed to
derive a mean-field theory. 
 Their model also involves  a scalar field $\psi$ coupled to both
an external electromagnetic field $A_{\mu}^{ext}$ and to a ``statistical'' gauge field $A_{\mu}$. It also includes a
potential term, and is time-dependent. Their Lagrangian reads
\begin{equation}
\begin{array}{ll}
\L_{ZHK}\;=%\;&4\theta\,\epsilon^{ij}\big(2A_0\p_{i}A_{j}-A_{i}\p_{0}A_{j}\big)
&-\frac{1}{4\theta}\epsilon^{\mu\nu\sigma}A_{\mu}\p_{\nu}A_{\sigma}
%\nonumber
\\[8pt]
&
+\psi^*\big[i\p_{t}-(A_{t}+A_{t}^{ext})\big]\psi
+\psi^*\big[-i\vnabla-(\vec{A}+\vec{A}^{ext})\big]^{2}\psi
+U(\psi),
\end{array}
\label{ZHKLag}
\end{equation}
where $A_{\mu}^{ext}$ is the vector potential of
an external electromagnetic field and
\begin{equation}
U(\psi)=\mu\vert\psi\vert^2-\lambda\vert\psi\vert^{4}
\label{ZHKpot}
\end{equation}
is a self-interaction potential. The term 
$\mu\vert\psi\vert^2$ ($\mu\geq0$) here is a
chemical potential, while the quartic term is an effective interaction
coming from the non-local expression
$$
\2\int \psi^\star(\vec{x})\psi^\star(\vec{x}')
V(\vec{x}-\vec{x}')
\psi(\vec{x})\psi(\vec{x}')d^2\vec{x}d^2\vec{x}'
$$
in the second-quantized Hamiltonian, when the two-body potential 
 is approximated by a delta function, 
$$ 
V(\vec{x}-\vec{x}')=-2\lambda\,\delta(\vec{x}-\vec{x}').
$$

For a static system in a purely magnetic 
background and for $U(\psi)\equiv0$, the two models are 
 mathematically equivalent, though \cite{HHY}.

Let us point out that the ZHK Lagrangian is \textit{first-order
in the time derivative} of the scalar field. It is
indeed non-relativistic, as will be shown in Section
\ref{background}.

Zhang et al argue that their model admits vortex-type solutions
\cite{ZHK,ZhangRev}, studied in \cite{Tafel} in some
detail.  

%%%%%%%%%%%%%%%%%%%%%%%%%%%%%%%%%%%%%%%%%%%%%%
\section{Relativistic Chern-Simons vortices}\label{RVort}
%%%%%%%%%%%%%%%%%%%%%%%%%%%%%%%%%%%%%%%%%%%%%%

Instead of pursuing the evolution in condensed matter physics, now we turn to study the field-theoretical aspects.

The first (Abelian)\footnote{A non-Abelian  theory
with vortex solutions has also  been proposed, cf.
\cite{NAbCS, DunneLN}.}  Chern-Simons model  
is obtained by simply adding the Chern-Simons term to the
usual Abelian Higgs model  \cite{PaKh}~:
\begin{eqnarray}
L=&\frac{1}{2}(D_\alpha\psi)^*D^\alpha\psi-
U(\psi)
-\frac{1}{4}F_{\alpha\beta}F^{\alpha\beta}
+
\frac{\kappa}{4}e^{\alpha\beta\gamma}A_\alpha F_{\beta\gamma},
\label{CSLag}
\\[8pt]
&U(\psi)=\frac{\lambda}{2}\big(1-|\psi|^2)^2,
\label{quarticpot}
\end{eqnarray}
where
$
D_\alpha\psi=\partial_\alpha\psi-ieA_\alpha\psi
$
is the covariant derivative, $e$  the 
electric charge of the field $\psi$. and
The Chern-Simons term is coupled through the
coupling constant $\kappa$. The theory lives in
$(2+1)$-dimensional Minkowski space, with the metric 
$(g_{\mu\nu})={\rm diag}\,(1,-1,-1)$, the coordinates being 
$x^0=t$ and $(x_i)=\vec{x}$.

The system can be studied along the same lines
as in the Nielsen-Olesen case \cite{NiOl}. Paul and Khare
\cite{PaKh} argue in fact that for the generalization to
$A_0\neq0$ of the Nielsen-Olesen radial Ansatz 
\begin{equation}
A_0=A_0(r),
\qquad
A_r=0,
\qquad
A_\vartheta=-\frac{A(r)}{r},
\qquad
\psi(r)=f(r)e^{-in\vartheta},
\label{PKradAns}
\end{equation}
the equations of motion,  
\begin{equation}
\begin{array}{l}
\displaystyle\frac{d^2A}{dr^2}
-\displaystyle\frac{1}{r}\frac{dA}{dr}
-ef^2(n+eA)
=\kappa r\frac{dA_0}{dr},
\\[12pt]
\displaystyle\frac{d^2A_0}{dr^2}
+\displaystyle\frac{1}{r}\frac{dA_0}{dr}
-e^2A_0f^2=\kappa\,\displaystyle\frac{1}{r}
\frac{dA}{dr},
\\[14pt]
\displaystyle\frac{d^2f}{dr^2}
+\displaystyle\frac{1}{r}\displaystyle\frac{df}{dr}
-\displaystyle\frac{1}{r^2}\left(n+eA\right)^2f+e^2A_0^2f
=-4\lambda f(1-f^2),
\end{array}
\label{PKradsymEq}
\end{equation}
supplemented with the
 finite-energy asymptotic conditions
\begin{equation}\begin{array}{lllllllll}
\lim_{r\to\infty}f(r)&=&1,\qquad
&\lim_{r\to\infty}A(r)&=&-\displaystyle\frac{n}{e},\qquad
&\lim_{r\to\infty}A_0(r)&=&0
\\[8pt]
\lim_{r\to0}f(r)&=&0,
&\lim_{r\to0}A(r)&=&0,
&\lim_{r\to0}A(r)&=&0
\end{array}
\label{PKfinEn}
\end{equation}
will admit a solution for each integer $n$.
By (\ref{PKfinEn})
these solutions 
represent \textit{charged topological vortices} sitting at the origin, since they
 carry both \textit{quantized magnetic flux and electric charges}, 
\begin{equation}
\Phi=\frac{2\pi}{e}\,n,\quad
\qquad
Q=\kappa\,\frac{2\pi}{e}\,n=\kappa\,\Phi.
\label{PKFluxCharge}
\end{equation}
respectively. 

While these vortices have interesting physical properties,
the model suffers from the mathematical  difficulty of having
to solve second-order field equations.
Further insight can be gained if we turn off the Maxwell term
altogether, and trading the the standard fourth-order self-interaction scalar potential (\ref{quarticpot}) for a  $6^{th}$ order one,
\begin{equation}
U(\psi)=\frac{\lambda}{4}|\psi|^2\big(|\psi|^2-1\big)^2.
\label{6pot}
\end{equation}
The Euler-Lagrange equations read
\begin{eqnarray}
\2D_\mu D^\mu\psi&=&-\displaystyle\frac{\delta U}{\delta\psi^*}
\equiv-\displaystyle\frac{\lambda}{4}(|\psi|^2-1)\big(3|\psi|^2-1\big)\psi,
\label{relCSEL1}
\\[8pt]
\2\kappa\,\epsilon^{\mu\alpha\beta}F_{\alpha\beta}&=&ej^\mu,
\label{relCSEL2}
\end{eqnarray}
where
$
F_{\alpha\beta}=\partial_\alpha A_\beta-\partial_\beta A_\alpha
$ is the `electromagnetic' field,
and
$j^\mu\equiv(\varrho,\vec{\jmath})$ is the current
\begin{equation}
j^\mu=\frac{1}{2i}\big[\psi^*D^\mu\psi-\psi(D^\mu\psi)^*\big].
\label{current}
\end{equation}

The first of the equations  (\ref{relCSEL1}) is a nonlinear
Klein-Gordon equation (NLKG), familiar from the Abelian Higgs model \cite{NiOl}; 
the second, (\ref{relCSEL2}),
called the {\it Field-Curent Identity} (FCI), 
replaces the Maxwell equations. Let us observe that,
unlike the latter, these equations are of the first order
in the vector potentital.

It follows from the Bianchi identity,
that the current (\ref{current}) is conserved, 
\begin{equation}
\epsilon^{\alpha\beta\gamma}\,
\partial_\alpha F_{\beta\gamma}=0,
\quad\Rightarrow\quad
\partial_\mu j^\mu=0.
\label{Bianchi}
\end{equation}

\goodbreak
%%%%%%%%%%%%%%%%%%%%%%%%%%%%%%%%%%%%%%%%%%%%
%\subsubsection{Finite-energy configurations}
%%%%%%%%%%%%%%%%%%%%%%%%%%%%%%%%%%%%%%%%%%%%
%%%%%%%%%%%%%%%%%%%%%%%%%%%%%%%%%%%%%%%%%%%%
\subsection{Finite-energy configurations}
%%%%%%%%%%%%%%%%%%%%%%%%%%%%%%%%%%%%%%%%%%%%

Let us consider a static field configuration $(A_\mu,\psi)$.
The energy, defined as the space integral of the time-time
component of the
energy-momentum tensor associated with the Lagrangian, is
\begin{equation}
E\equiv\int\!d^2\vec{x}\,T^{00}
=\int\!d^2\vec{x}\Big[\2D_i\psi(D^i\psi)^*
-\2e^2A_0^2|\psi|^2+\kappa A_0B+U(\psi)\Big],
\label{relEn}
\end{equation}
where
$B=-F^{12}$
is the magnetic field. Note that this expression is {\it not}
positive definite.
Observe, however, that the static solutions of the equations of motion (\ref{relCSEL1})-(\ref{relCSEL2})
are  stationary points of the energy. 

Variation of (\ref{relEn}) w. r. t. $A_0$ yields one of the equations 
of motion, namely
\begin{equation}
-e^2A_0|\psi|^2+\kappa B=0.
\label{relconstr}
\end{equation}
Eliminating $A_0$ from (\ref{relEn}) using this equation, 
we obtain the {\it positive definite} energy functional
\begin{equation}
E=\int\!d^2\vec{r}\,\Big[\2D_i\psi(D^i\psi)^*
+\frac{\kappa^2}{2e^2}\,\frac{B^2}{|\psi|^2}
+U(\psi)\Big].
\label{relEnbis}
\end{equation}

We are interested in static, finite-energy configurations.
Finite energy at infinity is guaranteed by the conditions\footnote{These conditions are  
in no way necessary; they yield 
the so-called {\it topological solitons}. Non-topological solutions 
are constructed in Ref. \cite{JaLeWe}.}
\begin{equation}
\left\{\begin{array}{llll}
i.)&\qquad\vert\psi\vert^2-1
&\quad=&\quad{\rm o}\big({1/r}\big),
\\[8pt]
ii.)
&\qquad B&\quad=&\quad{\rm o}\big({1/r}\big),
\\[8pt]
iii.)&\qquad\vec{D}\psi
&\quad=&\quad{\rm o}\big({1/r}\big).
\end{array}\right.\qquad
{r\to\infty}.
\label{relfenercond}
\end{equation}
Therefore, the $U(1)$ gauge symmetry is broken  for large $r$.
In particular, the scalar field $\psi$ is
covariantly constant, $\vec{D}\psi=0$.
This equation is solved  by parallel transport,
\begin{equation}
\psi(\vec{x})=\exp\Big[i\int_{\vec{x}_0}^{\vec{x}}eA_idx^i\Big]\,\psi_0,
\label{partrans}
\end{equation}
which is well-defined whenever
\begin{equation}
\oint eA_idx^i=
\int_{\IR^2}\!d^2\vec{x}\,eB
\equiv e\Phi
=2\pi n,
\qquad
n=0,\pm1,\ldots.
\label{fluxquant}
\end{equation}
Thus, {\it the magnetic flux is quantized}.

By i.),
the asymptotic values of the Higgs field
provide us with a mapping from
the circle at infinity
into the vacuum manifold, which is again a circle, 
$|\psi|^2=1$. 
Since the vector potential behaves asymptotically as
\begin{equation}
A_j\simeq-\frac{i}{ e}\,\partial_j\log\psi,
\label{freegauge}
\end{equation}
the integer $n$ in Eq. (\ref{fluxquant}) is the
{\it winding number} of this mapping; it is also
called the
{\it topological charge} (or {\it vortex number}).

Spontaneous symmetry breaking generates mass \cite{DeYa}.
Expanding $j^\mu$ around the vacuum expectation value of $\psi$ 
we find 
$j^\mu=-eA^\mu$ so that (\ref{relCSEL2}) is approximately 
$$
\2\kappa\,\epsilon^{\mu\alpha\beta}F_{\alpha\beta}\approx-e^2A^\mu.
\quad\hbox{Hence}\quad
F_{\alpha\mu}\approx
-({e^2/\kappa})\,\epsilon_{\alpha\mu\beta}A^\beta.
$$
Inserting here $F_{\alpha\beta}$ and deriving by
 $\partial^\alpha$,
we find that the gauge field $A^\mu$
satisfies the Klein-Gordon equation
$$%\begin{equation}
\dAlembert A^\mu
\approx
-\Big(\frac{e^2}{\kappa}\Big)^2A^\mu,
%\label{gaugeKG}
$$%\end{equation}
showing that the mass of the gauge field is
\begin{equation}
m_A=\frac{e^2}{\kappa}.
\label{gfmass}
\end{equation}

The Higgs mass is found in turn expanding $\psi$ around its
expectation value, chosen
as $\psi_0=(1,0)$, 
 $(\psi_r,\psi_\vartheta)=(1+\varphi,\theta)$,
yielding
\begin{equation}
U=\underbrace{U(1)}_{=0}
\quad+\quad
\underbrace{
\frac{\delta U}{\delta|\psi|}\Big\vert_{|\psi|=1}}_{=0}\,\varphi
\quad+\quad
\frac{1}{2}\underbrace{
\left(\frac{\delta^2U}{\delta|\psi|^2}\right)
\Big\vert_{|\psi|=1}}_{m_\psi^2}
\varphi^2,
\label{potexp}
\end{equation}
since $|\psi|=1$ is a critical point of $U$. We conclude that
the mass of the Higgs particle is \footnote{This can also be seen by considering the radial equation (\ref{relradsymeq})
below.}.
\begin{equation}
m_\psi^2=\frac{\delta^2U}{\delta|\psi|^2}\Big\vert_{|\psi|=1}
={2\lambda}.
\label{Higgsmass}
\end{equation}

\goodbreak
%%%%%%%%%%%%%%%%%%%%%%%%%%%%%%%%%%%%%
%\subsubsection{Radially symmetric solutions \cite{RadsymRel}}
%%%%%%%%%%%%%%%%%%%%%%%%%%%%%%%%%%%%%%
%%%%%%%%%%%%%%%%%%%%%%%%%%%%%%%%%%%%%
\subsection{Radially symmetric solutions}
%%%%%%%%%%%%%%%%%%%%%%%%%%%%%%%%%%%%%%

For the radially symmetric Ansatz
\begin{equation}
A_0=A_0(r),
\qquad
A_r=0,
\qquad
A_\vartheta=A(r),
\qquad
\psi(r)=f(r)e^{-in\vartheta},
\label{radsymAns}
\end{equation}
the equations of motion (\ref{relCSEL1})-(\ref{relCSEL2}) read
\begin{equation}
\begin{array}{l}
\displaystyle\frac{1}{r}
\frac{dA}{dr}+\displaystyle\frac{e^2}{\kappa}f^2A_0=0,
\\[12pt]
r\displaystyle\frac{dA_0}{dr}+\displaystyle\frac{e^2}{\kappa}
f^2\left(\displaystyle\frac{n}{e}+A\right)=0,
\\[14pt]
\displaystyle\frac{d^2f}{dr^2}+\displaystyle\frac{1}{r}
\displaystyle\frac{df}{dr}
-\displaystyle\frac{e^2}{r^2}\left(\displaystyle\frac{n}{e}+A\right)^2f+e^2A_0^2f
=-\displaystyle\frac{\lambda}{4}f(1-f^2)(1-3f^2),
\end{array}
\label{relradsymeq}
\end{equation}
with asymptotic conditions
\begin{equation}\begin{array}{llllll}
\lim_{r\to\infty}A(r)\qquad&=&-\displaystyle\frac{n}{e},\qquad
\qquad
&\lim_{r\to\infty}f(r)&=&1,
\\[8pt]
\lim_{r\to0}A(r)&=&\ 0,
&\lim_{r\to0}f(r)&=&0.
\end{array}
\label{relradsymasymp}
\end{equation}
This is either seen by a direct substitution into the equations,
or by re-writing the energy as
\begin{equation}
E=2\pi\int_0^\infty dr\left\{
\frac{r}{2}\,(f')^2
+\frac{a^2}{ 2r}f^2+\frac{\kappa^2}{2e^4f^2}
\frac{(a')^2}{ r}
+U(f)\right\},
\label{relradsymen}
\end{equation}
where $a=eA+n$.
The upper equation in (\ref{relradsymeq}) is plainly the 
radial form of (\ref{relconstr}).
Then variation of (\ref{relradsymen}) with respect to $a$ 
and $f$ yields
the two other equations in (\ref{relradsymeq}).

Approximate solutions can be obtained by inserting the asymptotic value,
$f\approx1$, into the first two equations: 
\begin{equation}
\frac{a'}{r}+\frac{e^2}{\kappa}A_0=0,
\qquad
A_0'+\frac{e^2}{\kappa}\frac{a}{r}=0,
\label{relappeq1}
\end{equation}
from which we infer that
\begin{equation}
\frac{d^2A_0}{d\rho^2}
+
\frac{1}{\rho}\frac{dA_0}{d\rho}-A_0=0,
\qquad
\rho\equiv(e^2/\kappa)r.
\label{relappeq2}
\end{equation}
This is the modified Bessel equation [Bessel equation of imaginary 
argument] of order zero. Hence
\begin{equation}
A_0=CK_0(\frac{e^2}{\kappa}r).
\label{A_0Bess}
\end{equation}
Similarly, for $a={n/e}+A$ we find, 
putting $\alpha=a/r$, 
\begin{equation}
\alpha''+\frac{\alpha'}{\rho}
-\Big(1+\frac{1}{\rho^2}\Big)\alpha=0,
\label{alphaeq}
\end{equation}
which is Bessel's equation of order $1$ with imaginary argument.
Thus $\alpha=CK_1(\rho)$ so that
\begin{equation}
A=-\frac{n}{e}+C\frac{e^2}{\kappa}r\,K_1
\big(\frac{e^2}{\kappa}\,r\big).
\label{AeqBess}
\end{equation}

Another way of deriving this result is to express $A$ 
from the middle equation in (\ref{relradsymeq}),
\begin{equation}
A=-\frac{n}{ e}-\frac{\kappa}{ e^2}\,r\frac{\ d}{ dr}A_0.
\end{equation}
The consistency with (\ref{AeqBess}) follows from the recursion relation
$
K_0'=-K_1
$
of the Bessel functions. 

An even coarser approximation is
obtained by eliminating the $\frac{a'}{ r}$ 
term by setting $a=ur^{-1/2}$ and
dropping the terms with inverse powers of $r$. Then both 
equations reduce to
\begin{equation}
u''=\big(\frac{e^2}{\kappa}\big)^2u
\qquad\Longrightarrow\qquad
A_0=a=\frac{C}{\sqrt{r}}\,e^{-m_Ar},
\label{ueq}
\end{equation}
which shows that the fields approach their asymptotic values
exponentially, with characteristic length determined by the
gauge field mass.

The deviation of $f$ from its asymptotic value, $\varphi=1-f$, is
found by inserting $\varphi$ into the last eqn. of (\ref{relradsymeq});
developping to first order in $\varphi$ we get
\begin{equation}
\varphi''+\frac{1}{ r}\varphi'-2\lambda\varphi\simeq0
\qquad\Longrightarrow\qquad
\varphi=CK_0(\sqrt{2\lambda}\,r),
\label{phieq}
\end{equation}
whose asymptotic behaviour is again exponential
 with characteristic length
$(m_\psi)^{-1}$,
\begin{equation}
\varphi=\frac{C}{\sqrt{r}}\,e^{-m_\psi r}.
\label{phival}
\end{equation}
The penetration depths of the 
gauge and scalar fields are therefore 
\begin{equation}
\eta=\frac{1}{ m_A}=\frac{e^2}{\kappa}
\qquad\hbox{and}\qquad
\xi=\frac{1}{ m_\psi}=\frac{1}{\sqrt{2\lambda}},
\label{penet}
\end{equation}
respectively.
For small $r$ instead, 
inserting the developments in powers of $r$, we find
\begin{equation}
\begin{array}{l}
f(r)\sim f_0r^{|n|}+\ldots,
\\[12pt]
A_0\sim\alpha_0-\displaystyle\frac{ef_0^2n}{2\kappa|n|}\,r^{2|n|}+\ldots,
\\[12pt]
A\sim-\displaystyle\frac{e^2f_0^2\alpha_0}{2\kappa(|n|+1)}\,r^{2|n|+2}+\ldots,
\end{array}
\label{fA0Aexp}
\end{equation}
where $\alpha_0$ and $f_0$ are constants. 
In summary, 
\begin{equation}\begin{array}{ll}
|\psi(r)|&\equiv f(r)
\qquad\propto\qquad
\left\{\begin{array}{ccc}
r^{|n|}&\qquad r\sim0
\\[8pt]
1-Cr^{-1/2}\,e^{-m_\psi r}
\qquad&\qquad r\to\infty
\end{array}\right.
\\[20pt]
|E(r)|&=|A_0'(r)|
\quad\propto\qquad
\left\{\begin{array}{ccc}
r^{2|n|-1}&r\sim0
\\[8pt]
Cr^{-1/2}\,e^{-m_Ar}+\hbox{\small lower order terms}
\qquad&r\to\infty
\end{array}\right.
\\[20pt]
|B(r)|&=\frac{|A'|}{ r}
\qquad\propto\qquad
\left\{\begin{array}{ccc}
r^{2|n|}&r\sim0
\\[8pt]
Cr^{-3/2}\,e^{-m_Ar}+\hbox{\small lower order terms}
\qquad&r\to\infty
\end{array}\right.
\end{array}
\label{approxfield}
\end{equation}

\goodbreak
%%%%%%%%%%%%%%%%%%%%%%%%%%%%%%
\subsection{Self-dual vortices}
%%%%%%%%%%%%%%%%%%%%%%%%%%%%%%%
 
In the Abelian Higgs model, an important step
has been to recognize that, for a specific value of 
the coupling constant, the field equations could be reduced
to first-order \cite{Bogo,Taubes}. This can also be achieved
 by a suitable
modification of the model \cite{RelCSvort,JaLeWe}, we discuss
below in some detail.

Let us suppose that the fields have equal masses,
$m_\psi^2=m_A^2\equiv m^2$ and hence equal
penetration depths.
Then the Bogomolny trick applies, i.e., 
the energy can be rewritten in the form
\begin{equation}
E=\int d^2\vec{r}\,\left[
\2|(D_1\pm iD_2)\psi|^2
+\2\Big\vert\frac{\kappa}{ e}\,\frac{B}{\psi}
\mp\frac{e^2}{2\kappa}\psi^*(1-|\psi|^2)\Big\vert^2\right]
\mp\int d^2\vec{r}\,\frac{eB}{2}(1-|\psi|^2).
\label{relBogodec}
\end{equation}
The last term can also be presented as
$$ %\begin{equation}
\mp\frac{eB}{2}
\mp\2\vec{\nabla}\times\vec{\jmath}.
$$ %\end{equation}
The integrand of the $B$-term yields the magnetic flux; the second is
transformed, by Stokes' theorem, into the circulation of the current at
infinity which vanishes, since all fields drop off at infinity
by assumption.
Its integral is therefore proportional to the magnetic flux,
$
\pm{e\Phi/2}.
$
Since the first integral is non-negative, we
have, in conclusion,
\begin{equation}
E\geq \frac{e|\Phi|}{2}=\pi|n|,
\label{relBogobound}
\end{equation}
equality being only attained if the {\it self-duality equations}
\begin{eqnarray}
D_1\psi&=&\mp iD_2\psi
\label{relSD1}
\\[8pt]
eB&=&\pm\frac{m^2}{2}\,|\psi|^2(1-|\psi|^2)
\label{relSD2}
\end{eqnarray}
hold.
It is readily verified that the solutions of equations  
(\ref{relconstr}) and (\ref{relSD1}-\ref{relSD2}) solve
automatically the non-linear Klein-Gordon equation (\ref{NLKG}). 

Let us first study the radial case.
For the Ansatz (\ref{radsymAns}) the self-duality
equations become
\begin{equation}
f'=\pm\frac{a}{r}f,
\qquad
\frac{a'}{r}=\pm\2m^2f^2(f^2-1),
\label{relradSD}
\end{equation}
where we introduced again $a=eA+n$.
Deriving the first of these equations and using the 
second one, for $f$ we get the \textit{Liouville - type equation} 
\begin{equation}
\bigtriangleup\log f=\frac{m^2}{2}\,f^2(f^2-1).
\label{Lioutype}
\end{equation}
 
Another way of obtaining the first-order eqns. 
(\ref{relradSD}) is to rewrite, for 
\begin{equation}
U(f)=\frac{m^2}{8}\,f^2\big(f^2-1\big)^2,
\end{equation}
the energy as
\begin{equation}
\pi\!\int_0^\infty\!rdr\left\{
\big[f'\mp\frac{a}{ r}f\big]^2
+\frac{1}{ m^2f^2}\big[{a'}{ r}\mp\frac{m^2}{2}f^2(f^2-1)\big]^2
\right\}
\pm\pi (af^2)\Big\vert_0^\infty
\mp\pi a\Big\vert_0^\infty.
\end{equation}
The boundary conditions read
\begin{equation}
\begin{array}{cccc}
a(\infty)&=0,\qquad
&f(\infty)&=1,
\\[8pt]
a(0)&=n,
&f(0)&=0,
\end{array}
\label{bigsmallrbeh}
\end{equation}
and thus
$
E\geq\pi|n|
$
as before, with equality attained iff the equations (\ref{relradSD}) hold.

For $n=0$ the only solution is the vacuum,
\begin{equation}
f\equiv1,
\qquad
A\equiv1.
\label{relvacuumsol}
\end{equation} 
To see this, note that the boundary conditions at infinity are
$
f(\infty)=1
$
and
$
A(\infty)=0.
$
Let now $f(r)$, $A(r)$ denote an arbitrary finite-energy 
configuration and consider 
$$
f_\tau(r)=f(r),
\qquad
A_\tau(r)=\tau A(r)
$$
where $\tau>0$ is a real parameter. This provides us with a
$1$-parameter family configurations with finite energy
$$
E_\tau=2\pi\int_0^\infty dr\left\{
\frac{r}{2}\,\big(f'\big)^2
+\tau^2\bigg[\frac{a^2}{ 2r}f^2
+\frac{r}{2m^2}\big(\frac{a'}{ rf}\big)^2\bigg]
+U(f)\right\},
$$
which is a monotonic function of $\tau$, whose minimum
is at $\tau=0$ i.e. for $a\equiv0$. Then Eq. (\ref{relradSD}) implies that
$f'\equiv0$ so that $f\equiv1$ is the only possibility.

Let us assume henceforth that $n\neq0$.
No analytic solution has been found so far. 
To study the large-$r$ behaviour, put
$\varphi\equiv1-f$.
Inserting $f\approx1$, Eqs. (\ref{relradSD}) reduce to 
$$
\varphi'=\mp\frac{a}{ r},
\qquad
\frac{a'}{r}=\mp m^2\varphi.
$$
Deriving, we get
$$
\begin{array}{l}
\varphi''+\frac{1}{r}\varphi'-m^2\varphi=0
\quad\Longrightarrow\quad
\varphi=CK_0(mr),
\\[10pt]
a''-\frac{1}{r}a'-m^2a=0
\quad\Longrightarrow\quad
a=CmrK_1(mr).
\end{array}
$$
Thus, for large $r$,
\begin{equation}
f\approx1-CK_0(mr)
\and
A\approx-\frac{n}{e}+CmrK_1(mr)
\label{AsymprRelSol}
\end{equation}
with some constant $C$. 
For small $r$ instead, Eq. (\ref{relradSD}), yields, to 
${\rm O}(r^{5|n|+1})$, the expansion
\begin{equation}\begin{array}{l}
f(r)= f_0r^{|n|}
-\displaystyle\frac{f_0^3m^2}{2(2n+2)^2}\,r^{3|n|+2}
+{\rm O}(r^{5|n|+2}).
\\[16pt]
A=-\displaystyle\frac{f_0^2m^2}{2(2|n|+2)e}\,r^{2|n|+2}
+\displaystyle\frac{f_0^2m^2}{2(4|n|+2)e}\,r^{4n+2}
+{\rm O}(r^{4|n|+4}).
\end{array}
\label{relsmallrexp}
\end{equation}

The result is consistent with (\ref{fA0Aexp}) since the constant 
$\alpha_0$ is now
$
\alpha_0={m/2e}={e/2\kappa}.
$
\begin{figure}
\hskip-0.8truecm
\begin{center}
\includegraphics[scale=0.55]{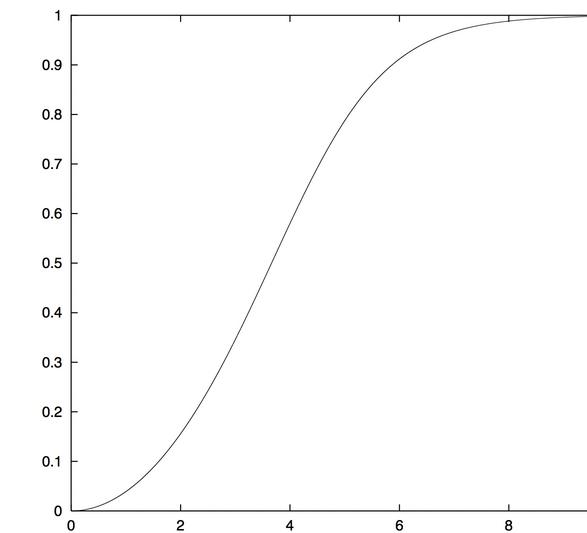}
\qquad
\includegraphics[scale=0.55]{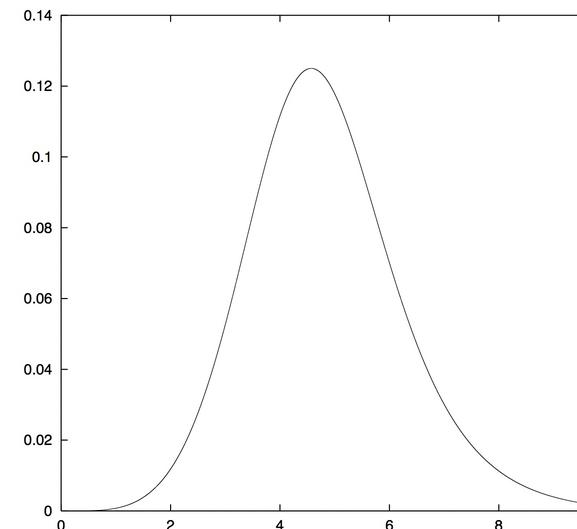}
\end{center}
\label{fp}
\caption{\it The scalar and the magnetic fields of the 
radially symmetric charge-$2$ relativistic vortex.
Note that $B=0$ where the scalar field vanishes, so that the magnetic field has a doughnut-like shape.}
\end{figure}

Let us mention that the asymptotic behaviour expressed in
Eq. (\ref{bigsmallrbeh}) is actually valid in full generality, without
the assumption of radial symmetry. 
Expressing in fact the vector-potential 
from the self-duality condition $(D_1\pm iD_2)\psi=0$ as
\begin{equation}
e\vec{A}=\vec\nabla({\rm Arg}\,\psi)\pm\vec\nabla\times\log|\psi|
\end{equation}
and inserting into the second equation in (\ref{relradSD}), we get 
again (\ref{bigsmallrbeh}), with $|\psi|$ replacing $f$.

Index-theoretical
calculations show that, for topological charge $n$,
Eqn. (\ref{relSD1}-\ref{relSD2}) admits a $2|n|$ parameter family of 
solutions \cite{RelCSvort}.

%%%%%%%%%%%%%%%%%%%%%%%%%%%%%%%%%%%
\section{Non-relativistic vortices}\label{NRVort}
%%%%%%%%%%%%%%%%%%%%%%%%%%%%%%%%%%%

The non-relativistic limit of the system studied in Section
\ref{RVort} is found \cite{JaPiRev,DunneLN} by
setting
\begin{equation}
\psi=e^{-imc^2t}\Psi+e^{+imc^2t}\bar\Psi,
\label{NRlim}
\end{equation}
where $\Psi$ and $\bar\Psi$ denote the particles and
antiparticles, respectively. Inserting (\ref{NRlim}) into the
action, dropping the oscillating terms
and only keeping those of order $1/c$,
shows that both the
particles and antiparticles are separately conserved. 
We can therefore consistently set $\bar\Psi=0$.
The remaining matter Lagrangian reads
\begin{equation}
\L_{{\small matter}}=
i\Psi^\star D_t\Psi
-\frac{|\vec{D}\Psi|^2}{2m}
+
\frac{\Lambda}{2}
(\Psi^\star\Psi)^2,
\label{NRCSLag}
\end{equation}
where $\Lambda={e^2/mc|\kappa|}$. At first, we will
let the constant $\Lambda$ be arbitrary.

It will be shown in Section \ref{NRSymmetries} below
that the theory is  non-relativistic see.
 
In what follows, we put $c=1$.

Variation of $\displaystyle\int\L_{matter}$ w. r. t. $\Psi^\star$ yields the
{\it gauged non-linear Schr\"odinger equation}
\begin{equation}
i\partial_t\Psi=
\left[
-\frac{\vec{D}^2}{2m}
-eA_t
-\Lambda\Psi^\star\Psi\right]\Psi,
\label{NLS}
\end{equation}
where $\vD=\vnabla-ie\vA$.
A self-consistent system is obtained by
adding the matter action to the Chern-Simons action (\ref{CSform}).
The variational 
equations are the Chern-Simons equations (\ref{relCSEL2})
written in non-relativistic notations, 
\begin{eqnarray}
B&\equiv\epsilon^{ij}\partial_iA^j=-\displaystyle\frac{e}{\kappa}\,\varrho,
&\hbox{Gauss}
\label{Gauss}
\\[6pt]
E^i&\equiv-\partial_iA^0-\partial_tA^i=
\displaystyle\frac{e}{\kappa}\epsilon^{ij}J^j,\qquad
&\hbox{FCI}
\label{FCI}
\end{eqnarray}
where 
\begin{eqnarray}
\varrho=\Psi^*\Psi
%\label{NRdens}
%\\[6pt]
\quad\hbox{and}\quad
J^\mu\equiv(\varrho,\vJ)%\vec\jmath)
=(\Psi^\star\Psi,\frac{1}{2mi}\big[
\Psi^*\vec{D}\Psi-\Psi(\vec{D}\Psi)^*\big])
\label{NRcurrent}
\end{eqnarray}
are the density and the current, respectively. The invariance
of (\ref{NRCSLag}) w.r.t. global gauge transformations 
$\Psi\to e^{i\omega}\Psi$ implies the
 continuity equation 
\begin{equation}
\p_t\varrho+\vnabla\cdot\vJ=0.
\label{conteq}
\end{equation}
Eqns. (\ref{Gauss}) and (\ref{FCI}) are
called the Gauss' law and the field-current identity (FCI),
respectively. 

%%%%%%%%%%%%%%%%%%%%%%%%%%%%%%%%%%%%%%%%%
\subsection{Self-dual NR vortex solutions}
%%%%%%%%%%%%%%%%%%%%%%%%%%%%%%%%%%%%%%%%

We would again like to find static soliton solutions.
The construction of an energy-momentum tensor is now more subtle,
because the theory is non-relativistic. A conserved
energy-momentum tensor can, nevertheless, be constructed
 \cite{JaPi2,JaPiRev,DHP1,MaCha}, see Section \ref{NRSymmetries}
below. It provides us with the energy functional
\begin{equation}
E=
\int d^2{x}\,
\Big(\frac{|\vec{D}\Psi|^2}{2m}-\frac{\Lambda}{2}(\Psi^\star\Psi)^2\Big).
\label{NRenfunc}
\end{equation}
Now we apply once again the Bogomolny trick.
Using the identity
\begin{equation}
\vert\vD\Psi\vert^2=
\vert(D_1\pm iD_2)\Psi\vert^2
\pm m\vnabla\times\vJ\pm eB\,\varrho,
\label{identity1}
\end{equation}
the energy (\ref{NRenfunc}) is  written as
\begin{equation}
E=
\int d^2{x}\,\frac{\big|(D_1\pm iD_2)\Psi|^2}{2m}
-\2\left(\Lambda-\frac{e^2}{m|\kappa|}\right)
(\Psi^\star\Psi)^2.
\label{NRBog}
\end{equation}
The energy is, hence, positive definite if
\begin{equation}
\Lambda\leq\frac{e^2}{m|\kappa|},
\label{posdefLambda}
\end{equation}
that we assume henceforth. The vacuum is clearly
\begin{equation}
\vA=0,\qquad\Psi=0.
\label{NRvacuum}
\end{equation}
To get finite energy, the following large-$r$ asymptotic 
behaviour is required~:
\begin{eqnarray}\left\{\begin{array}{ll}
\vD\Psi\to0
\\[10pt]
|\Psi|\to0
\end{array}\right.
\qquad\hbox{as}\quad r\to\infty.
\label{finEnCond}
\end{eqnarray}
The second condition here implies
that the finite-energy vortices constructed below are \textit{non-topological}~:
$
\Psi|_\infty:S_\infty\to0.
$

For the  specific value \footnote{
(\ref{specLambda}) is the same as the one we obtained above by taking the
non-relativistic limit of the self-dual relativistic
theory.} 
\begin{equation}
\Lambda=\frac{e^2}{m|\kappa|}
\label{specLambda}
\end{equation}
of the non-linearity, in particular,
the second term vanishes. 
Then the absolute minimum of the energy, namely zero, is attained for 
{\it self-dual} or {\it antiself-dual fields}
\footnote{
The equations (\ref{NRSD}-\ref{SDGauss}) can indeed be derived,
by symmetry reduction, from the $4D$ self-dual
Yang-Mills  equations \cite{Grossman}.}, i.e. for such that
\begin{equation}
D_\pm\Psi=0,
\where
D_\pm=D_1\pm iD_2.
\label{NRSD}
\end{equation}

Do we get a static solution of the problem by minimizing
the energy~? Let us first observe that 
the energy functional
(\ref{NRenfunc}) does not include the time component,
$A_t$, which should be fixed by the field equations.
For the self-dual Ansatz (\ref{NRSD}), the current is 
expressed as
\begin{equation}
\vJ=\pm\frac{1}{2me}\vnabla\times\varrho.
\label{SDrho}
\end{equation}
Using another identity, namely
\begin{equation}
\vD^2\Psi=\left(D_+D_-+eB\right)\Psi,
\label{identity2}
\end{equation}
the static field equations can be written as
\begin{eqnarray}
\left[\displaystyle\frac{D_+D_-}{2m}+\left(\Lambda\mp
\displaystyle\frac{e^2}{2m\kappa}\right)\varrho
-eA_t\right]\Psi=0
\label{SDNLS}
\\[6pt]
\kappa B-e\varrho=0
\label{SDGauss}
\\[7pt]
\vnabla \left(A_t-\frac{1}{2m\kappa}\varrho\right)=0
\label{SDFCI}
\end{eqnarray}
By inspection, using
$
\big[D_+,D_-\big]=eB,
$
we infer
that a static solution is obtained,
for the specific value (\ref{specLambda}), 
for
\begin{eqnarray}
&D_\pm\Psi=0
\label{SDbis}
\\[6pt]
&\kappa B+e\Psi\Psi^*=0,
\label{Gaussbis}
\end{eqnarray}
when the time component of the potential is
\begin{equation}
A_t=\frac{1}{2m\kappa}\varrho.
\label{SDAt}
\end{equation} 

Separating the phase as $\Psi=e^{ie\omega}\sqrt\varrho$, the
SD equation is solved by
\begin{equation}
\vA=\frac{1}{2e}\vec\nabla\times\log\varrho+
\vec\nabla\omega.
\label{NRvpot}
\end{equation}
When we insert $B=\vnabla\times\vA$ into (\ref{Gaussbis})
 $\varrho$ has to solve the \textit{Liouville equation},
\begin{equation}\fbox{$
\bigtriangleup\log\varrho=\pm\displaystyle\frac{2e^2}{\kappa}\varrho,
$}
\label{Liouville}
\end{equation}
cf. (\ref{Lioutype}). 

Having solved this equation, the scalar and
vector potentials are given by (\ref{SDAt}) and
(\ref{NRvpot}), respectively. In the latter, the
phase $\omega$ has to be chosen so that it
cancels the singularity due to the zeros of $\varrho$.
This property is related to the quantization of
the vortex charge, see Section \ref{LiouVort} below. 
The vectorpotential will be given in (\ref{genvecpot})
below.

%%%%%%%%%%%%%%%%%%%%%%%%%%%%%%%%%%%%%%%%%%%%%%%%%%%%%%
\subsection{Vortex solutions of the Liouville equation}\label{LiouVort}
%%%%%%%%%%%%%%%%%%%%%%%%%%%%%%%%%%%%%%%%%%%%%%%%%%%%%%

The vortices  are hence constructed out of
the  solutions of the Liouville equation (\ref{Liouville}).
Solutions defined over the whole plane arise when the r.h.s. is negative.
Hence, 
the upper sign has to be chosen when $\kappa<0$ and
the lower sign when $\kappa>0$. Then the general solution
reads
\begin{equation}\fbox{$
\varrho=\displaystyle\frac{4|\kappa|}{e^2}\,
\displaystyle\frac{|f'|^2}{(1+|f|^2)^2},
$}
\label{genLiouSol}
\end{equation}
where $f$ is a meromorphic function of $z=x+iy$.
For the radial Ansatz
\begin{equation}
f(z)=z^{-N}
\label{radAns}
\end{equation}
we obtain, in particular, the radially symmetric
solution
\begin{equation}
%\varrho(r)=\frac{4N^2|\kappa|}{e^2r^2}
%\left[\frac{|c|}{r^N}+\frac{r^N}{|c|}\right]^{-2},
\varrho(r)=\frac{4N^2|\kappa|}{e^2}\,|\frac{r^{-2(N+1)}}{(1+r^{-2N})^2}.
\label{radrho}
\end{equation}

\begin{figure}
\begin{center}
\includegraphics[scale=0.9]{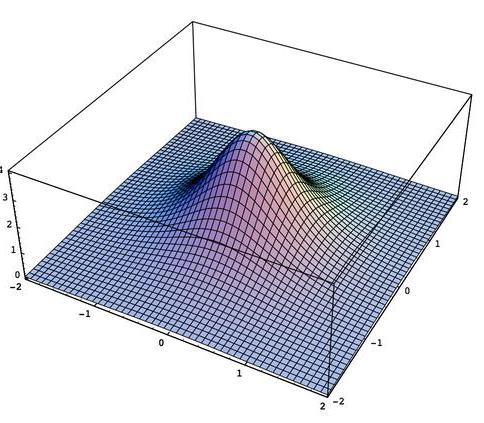}
\end{center}\vspace{-7mm}
\label{vortex1}
\caption{\it The non-relativistic radially symmetric $N=1$ vortex
has a maximum at $r=0$.}
\end{figure}

\begin{figure}
\begin{center}
\includegraphics[scale=0.8]{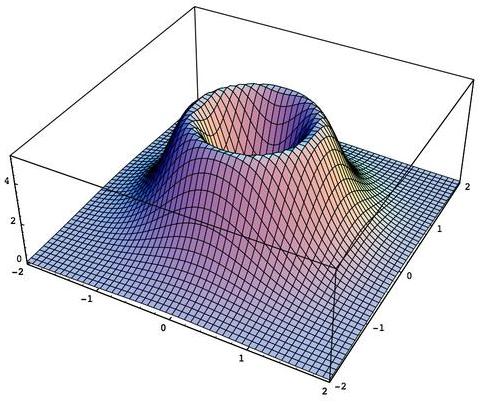}
\qquad
\includegraphics[scale=0.8]{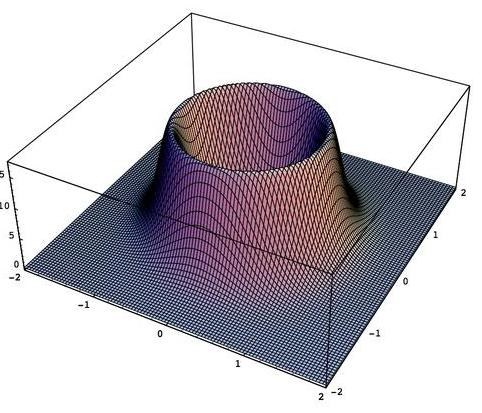}
\end{center}\vspace{-7mm}
\label{vortex2-4}
\caption{\it For $N\geq2$, the non-relativistic radially symmetric  
a vortices have a \lq doughnut-like' shape: the particle density
vanishes at $r=0$. The figure shows those  with $N=2$ and $N=4$.}
\end{figure}

The regularity requires, furthermore, that $N$ be an integer
at least $1$. For $N=1$ (Fig. \ref{vortex1}), the origin is
a maximum of $\varrho$; for $N\geq2$, it is a zero~:
the vortex has a ``doughnut-like'' shape, see 
Fig. \ref{vortex2-4}. Presenting (\ref{radrho}) as
\begin{equation}
\varrho(r)=\frac{4N^2|\kappa|}{e^2}
\left[\frac{r^{(N-1)}}{1+r^{2N}}\right]^2
\end{equation}

shows, forthermore, that the density $\varrho$ 
[and hence the magnetic field $B$] vanishes at the origin, 
$\varrho(0)=0=B(0)$, except for $N=1$, cf. the figures.
 Owing to the Gauss law (\ref{Gauss}),
the magnetic field behaves as in fact as
\begin{equation}
B\propto-\varrho\sim r^{2(N-1)}.
\end{equation}
The singularity in the first term in 
the vector potential $A$ (\ref{NRvpot}) can be canceled
choosing the phase of $\psi$ as 
\begin{equation}
\omega=(N-1)\theta,
\label{radvecpot}
\end{equation}
where $\theta$ is the polar angle of the position vector 
\cite{JaPi2}. 
 
Returning to the general case, we observe that not all meromorphic function yield
a physically interesting solution, though. 
The natural requirement is that
the magnetic and scalar fields, $B$ and $\Psi$, must be regular, 
and that the magnetic flux,
\begin{equation}
\Phi=\int\!B\,d^2x,
\label{vflux}
\end{equation}
be finite. For the radial Ansatz (\ref{radAns}) we find, for example,
\begin{equation}
%\Phi=2N ({\rm sign}\, \kappa)\Phi_0, 
\Phi=-\frac{4\pi\,N({\rm sg}\,\kappa)\hbar}{e}.
\end{equation}

Which functions $f$ yield regular, finite-flux solutions~? How can we calculate
the flux~? Is it quantized~? How many independent solutions do we get for a fixed value
$\Phi$~? The answers in \cite{CSflux,CSindex} are not entirely satisfactory: on the one hand, the proof given in
\cite{CSflux} is based on an asymptotic behaviour, that is only valid in the radial case.
On the other hand, the parameter-counting given in \cite{CSindex}
uses an index theorem, which is an unnecessary complication
here, when explicit solutions are known.
Elementary proofs were found in \cite{HY}. 

\kikezd{Theorem 1} \cite{HY}~: \textit{The meromorphic function $f(z)$ yields a regular vortex solution
with finite magnetic flux if and only if $f(z)$ is a rational function,
\begin{equation}
f(z)=\frac{P(z)}{Q(z)}
\qquad\hbox{\rm s.t.}\qquad {\rm deg}\, P< {\rm deg}\, Q,
\label{ratfunc}
\end{equation}
where the highest-order term on $Q$ can be normalized to $1$.}
\vspace{2mm}

In particular, when all roots of $Q(z)$ are simple,
$f(z)$ can be developed into partial fractions,
\begin{equation}
f(z)=\sum_{i=1}^N\frac{c_i}{z-z_i},
\end{equation}
where the $c_i$ and the $z_i$ are $2n$ complex numbers,
we get the $4N$-parameter family of 
$N$ separated one-vortices \cite{JaPi2}.
Note that this formula
breaks down for superimposed vortices.

The proof proceeds through a series of Lemmas \cite{HY}, and 
amounts to 
showing that $f$ can only have a finite number of isolated singularities that can not be essential
neither at a finite point, nor at infinity. Then a theorem 
of complex analysis \cite{Ahlfors} says that $f$ is necessarily rational. 
 
The density (\ref{genLiouSol}) is readily seen to be invariant 
w.r.t. 
\begin{equation}
f\to\frac{f+c}{1-\bar{c}f}.
\end{equation}
In particular, taking $c$ imaginary and letting
it go to infinity, it is invariant under changing 
$f$ into $1/f$. Hence ${\rm deg}\, P\leq {\rm deg}\, Q$ can be assumed.
But ${\rm deg}\, P={\rm deg}\, Q$ can be eliminated by a 
suitable redefinition  \cite{JaPi2}.

\kikezd{Theorem 2} \cite{HY}~: \textit{
The magnetic flux of the solution generated by $P/Q$
is evenly quantized,}
\begin{eqnarray}
\Phi=2N ({\rm sign}\, \kappa)\Phi_0,
\qquad
N={\rm deg}\, Q,
\qquad\Phi_0=-2\pi\frac{\hbar}{e}.
\label{NRfluxquant}
\end{eqnarray}

The proof amounts to showing that only the 
roots of the denominator 
\begin{equation}
Q(z)=(z-z_1)^{n_1}\dots(z-z_m)^{n_m},
\qquad(\sum_k n_k=N)
\end{equation}
contribute to the charge.
(\ref{NRfluxquant}) is inferred by transforming the
flux (\ref{vflux}) into a contour integral along the circle 
at infinity $C$. The isolated zeros of $Q(z)$,  $z_1,\dots,z_m$,
are identified with the ``positions'' of the
vortices.
Each of them can be encircled  by disjoint circles $C_k$,
and the charge comes form these zeros,
\begin{eqnarray}
\Phi=\oint_{C}=\sum_k\oint_{C_k}
=\sum_k n_k\left(-({\rm sg}\,\kappa)\frac{4\pi\hbar}{e}\right)=
-2N({\rm sg}\,\kappa)\Phi_0.
\end{eqnarray}
\begin{figure}
\begin{center}
\includegraphics[scale=0.4]{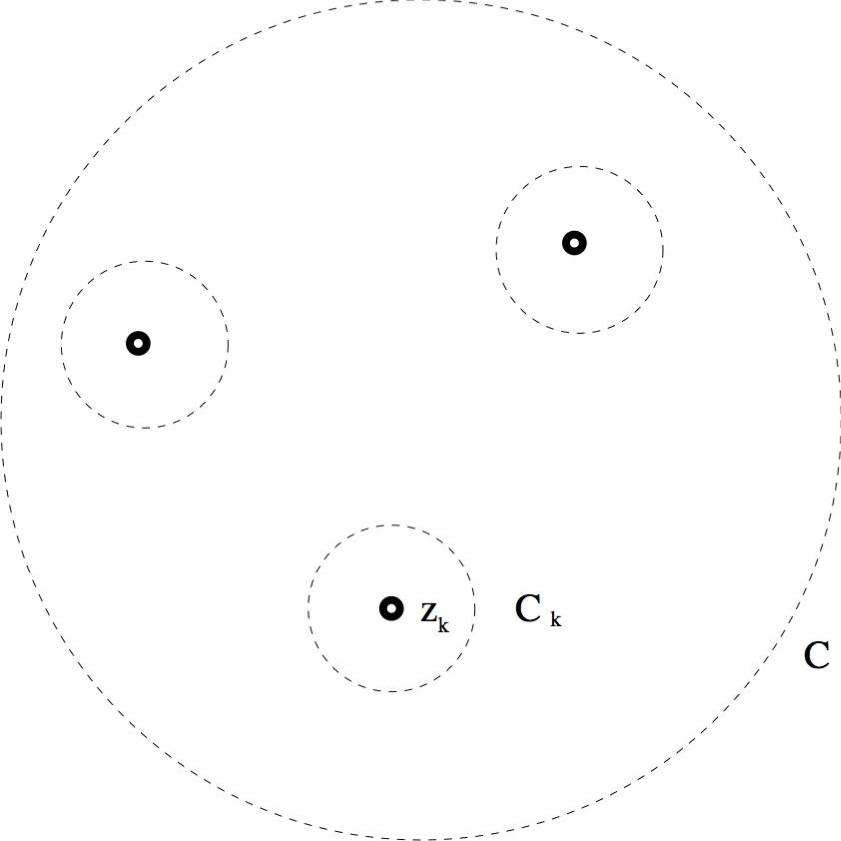}
\end{center}\vspace{-5mm}
\label{vcharge}
\caption{\it The charge of a vortex is proportional to the 
sum of the
multiplicities the zeros of the denominator $Q(z)$ in 
(\ref{ratfunc}).}
\end{figure}

Let us fix $N={\rm deg}\, Q> {\rm deg}\, P$.

\kikezd{Theorem 3.}~: \textit{
The solution generated by (\ref{ratfunc})  depends on $4N-1$ parameters,
 where $N$ is the degree of the denominator $Q(z)$}.
\vspace{3mm} 
 
The proof follows at once from Theorem 1.~: $N$ is the degree
of the denominator, $Q(z)$, which, being normalized, has 
$N$ complex coefficients. Due to deg $P<$ deg $Q$, the numerator
also has $N$ complex coefficients (some of which can vanish).
The $(-1)$, (missed in \cite{HY}) comes from noting that,
by (\ref{genLiouSol}), the general phase of $f$ is irrelevant, so that the highest coefficient of
$P$ can be chosen to be real.
\begin{figure}
\begin{center}
\includegraphics[scale=0.9]{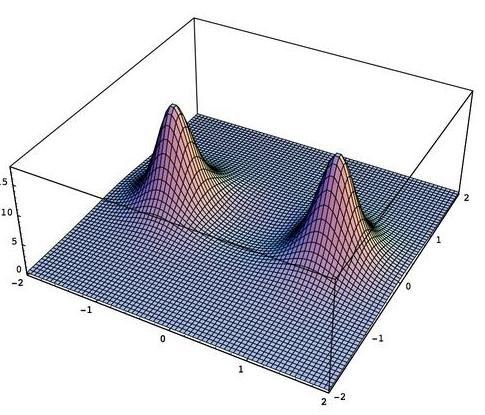}
\end{center}
\label{vortex3}\vspace{-7mm}
\caption{\it Two separated $1$-vortices with charge $N=2$.}
\end{figure}
\begin{figure}
\begin{center}\vspace{-20mm}
\includegraphics[scale=0.9]{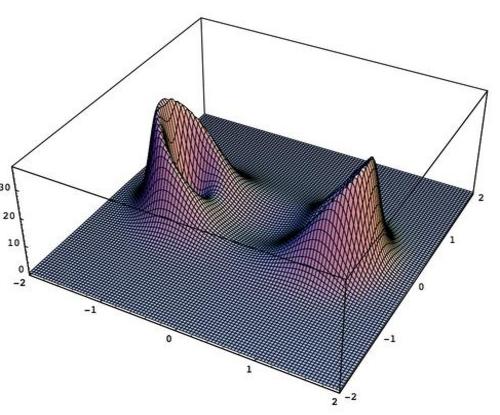}
\end{center}
\label{vortex5}\vspace{-7mm}
\caption{\it Two separated charge-2 vortices
with total charge $N=4$.}
\end{figure}

These results have a rather elegant
geometric interpretation \cite{topnontop}.
A rational function 
\begin{equation}
w=\frac{P(z)}{Q(z)}=\frac{a_mz^m+\dots+a_0}
{b_nz^n+\dots+b_0}
\label{ratfunc2}
\end{equation}
($a_m,b_n\neq0$)
 always has a limit
as $z\to\infty$, namely $\infty$ if 
$m={\rm deg}\,Q>{\rm deg}\,Q=n$, 
$a_m/b_n$ if $m=n$, and
zero, if $m<n$. It extends therefore as a mapping, 
still denoted by $f$, between the Riemann spheres, 
\begin{equation}
f~:S_z\to S_w,
\end{equation}
 obtained by compactifying the complex 
$z$ and $w$-planes by adding the point at infinity. Then
the $z$ and $w$ are stereographic coordinates. 
\begin{figure}
\begin{center}\vspace{-14mm}
\includegraphics[scale=.9]{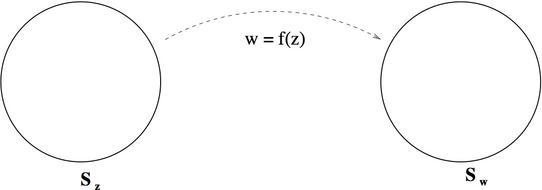}
\end{center}\vspace{-6mm}
\label{SpherTop}
\caption{\it The magnetic charge of a non-topological vortex
is in fact the topological charge in monopole theory.}
\end{figure}
The $w$-sphere
carries, in particular, the canonical surface form
\begin{equation}
\Omega=2i\frac{dw\wedge d\bar{w}}{(1+w\bar{w})^2}.
\label{surf}
\end{equation}
Using the Gauss law $B=-(\frac{e}{\kappa})\varrho$,
the magnetic flux of the vortex, $\Phi=\displaystyle\int Bd^2x$,  is \footnote{Remember that in our units
$2\pi\hbar=h=1$.}
\begin{eqnarray}
\Phi=
-({\rm sg}\,\kappa/e) \displaystyle\int
\frac{4|f'|^2}{(1+|f|^2)^2}d^2z
=
-(2/e)({\rm sg}\,\kappa)\displaystyle\int_{S_z}\!f^*\Omega,
\label{VortTopcharge}
\end{eqnarray}
where we recognize the \textit{topological charge}
of monopole theory \cite{monoptop}. The integral in
(\ref{VortTopcharge}) is in fact 
the same as the homotopy class of the mapping $f:~S_z\to S_w$.

Equivalently, the magnetic charge is the \textit{Brouwer degree} of $f$
[which is the number of times the image is covered].

Generalizing (\ref{radvecpot}),
 the regularity of the vector
potential requires chosing the phase $\omega$ so that \cite{HY} 
\begin{equation}
(\p_x-i\p_y)\omega=\sum_{i=1}^{N_Q}\frac{n_i-1}{z-z_i}+
\sum_{k=1}^{N_P}\frac{n_k+1}{z-Z_k}.
\label{genvecpot}
\end{equation}
where the $z_i, \,i=1,\dots n_Q$ are the distinct roots of the denominator
$Q(z)$ and  $n_i$ their respective multiplicity,  so that
$\sum_{i=1}^{N_Q}n_i=\hbox{deg}\,Q=N$ is the vortex number. The
$Z_k;\ k=1,\dots N_P$, are the roots of the numerator; their
multiplicities are $m_k$, and 
$\sum_{k=1}^{N_P}m_k=\hbox{deg}\,P<N$.

Remarkably, the self-dual solutions
of the $\O(3)$ non-linear sigma model (\cite{BePo}) are,
once again, precisely those described here.  

%%%%%%%%%%%%%%%%%%%%%%%%%%%%%%%%%%%%%%%%%%%%%%%%%%%
\subsection{Symmetries of non-relativistic vortices}\label{NRSymmetries}
%%%%%%%%%%%%%%%%%%%%%%%%%%%%%%%%%%%

A subtle point of non-relativistic CS theory
is the construction of a conserved energy-momentum
tensor. Jackiw and Pi \cite{JaPi2} present the rather
complicated-looking expressions
\begin{equation}
\begin{array}{lll}
T^{00}&=&\frac{1}{2m}|\vD\Psi|^2-\frac{\Lambda}{2}|\Psi]^4,
\\[10pt]
T^{i0}&=&-\frac{1}{2}\left(
(\vD_t\Psi)^*(D_i\Psi)+(D_i\Psi)^*D_t\Psi\right),
\\[10pt]
T^{0i}&=&
-\frac{i}{2}\left(\Psi^*D_i\Psi-(D_i\Psi)^*\Psi\right),
\\[10pt]
T^{ij}&=&-\frac{1}{2}\left(
(D_i\Psi)^*D_j\Psi+(D_j\Psi)^*D_i\Psi
-\delta_{ij}\,|\vD\Psi|^2\right)
\\[10pt]\qquad
&&+\frac{1}{4}\left(\delta_{ij}\,\bigtriangleup
-2\p_i\p_j\right)(|\Psi|^2)+\delta_{ij}\,T^{00},
\end{array}
\label{Tmunu}
\end{equation}
whose conservation,
\begin{equation}
\p_\alpha T^{\alpha\beta}=0,
\end{equation}
can be checked by a diract calculation.
The tensor $T^{\alpha\beta}$ is, however, symmetric only in the spatial indices, 
\begin{equation}
T^{0i}\neq T^{i0},
\qquad
T^{ij}=T^{ji}.
\label{NRsymEM}
\end{equation}
$T^{ij}$ has been ``improved''
and satisfies, instead of the usual
tracelessness-condition $T^{\alpha}_{\ \alpha}=0$
of relativistic field theory, the modified trace condition
\begin{equation}
T^{i}_{\ i}=2T^{00}.
\label{tracecond}
\end{equation} 
These unusual properties are, as we explain it below,
hallmarks of Schr\"odinger, rather then 
Lorentz-conformal invariance.

Let us remind the Reader the definition~: 
a {\it symmetry} is a transformation 
which interchanges the solutions of the coupled equations of motion.
For a Lagrangian system, an infinitesimal space-time symmetry
can be represented
 by a vector field $X^\mu$ on space-time,
is a symmetry, when it changes the Lagrangian by a surface term,
\begin{equation}
\L\to\L+\partial_{\alpha}K^\alpha
\end{equation}
for some function $K$.  
To each such transformation, N{\oe}ther's theorem
associates a conserved quantity, namely
\begin{equation}
C=\int\left(\frac{\delta\L}{\delta(\partial_{t}\chi)}
\delta\chi-K^t\right)d^2\vec{x},
\end{equation}
where $\chi$ denotes, collectively, all fields.

The Galilean symmetry of our Chern-Simons-theory 
follows from
the general framework  \cite{Hagen}. To
each generator of the centrally-extended Galilei group
is associated a conserved quantity, namely
\begin{equation}
\begin{array}{ll}
\cH=\displaystyle\int T^{00}d^2x,&\hbox{energy}
\\[10pt]
\cP_i= \displaystyle\int T^{0i}\,d^2x,
\qquad
&\hbox{momentum}
\\[10pt]
\cJ= \displaystyle\int \epsilon_{ij}x^iT^{0j},\, d^2x&\hbox{angular momentum}
\\[10pt]
\cG_i=t\cP_i-m \displaystyle\int x_i\varrho d^2x
\qquad\quad&\hbox{center of mass}
\\[10pt]
\cN=m \displaystyle\int x_i\varrho d^2x
&\hbox{mass (particle number)}
\end{array}
\end{equation}
What is less expected is that the model admits 
 two more conserved generators, namely
\begin{equation}
\begin{array}{ll}
\cD=t\cH-\frac{1}{2}x_i\cP_i
&\hbox{dilatation}
\\[10pt]
\cK=-t^2\cH+2t\cD+\displaystyle\frac{m}{2}\int r^2\varrho\, d^2x
\qquad\quad
&\hbox{expansion}
\end{array}
\label{SchrGen}
\end{equation}
 
The Poisson brackets,
\begin{equation}
\Big\{f,g\Big\}=\int\sum_i\left(
\frac{\p f}{\p x_i}\frac{\p g}{\p p_i}
-
\frac{\p g}{\p x_i}\frac{\p f}{\p p_i}
\right) d^2x
\label{PBrackets}
\end{equation}
 of these conserved quantities 
are those which define the  non-relativistic ``conformal'' extension of the
Galilei group, called the Schr\"odinger group \cite{SchrG}
\begin{equation}
\begin{array}{llllllllll}
\big\{{\cal G}_i,{\cal G}_j\big\}\hfill&=&0,\hfill
&\big\{{\cal P}_i,{\cal P}_j\big\}\hfill&=&0,\hfill
&\big\{{\cal P}_i,{\cal G}_{j}\big\}\hfill&=&\delta_{ij}\,\cN,\hfill
\\[8pt]
\big\{{\cal G}_i,{\cal R}\big\}\hfill&=&\epsilon_{ij}\,{\cal G}_{j},\qquad\hfill
&\big\{{\cal P}_i,{\cal R}\big\}\hfill&=&\epsilon_{ij}{\cal P}_{j},\qquad\hfill
&&&
\\[8pt]
\big\{{\cal H},{\cal G}_i\big\}\hfill&=&{\cal P}_i,\hfill
&\big\{{\cal H},{\cal P}_i\big\}\hfill&=&0,\hfill
&\big\{{\cal H},{\cal R}\big\}\hfill&=&0,\hfill
\\[8pt]
\big\{{\cal H}, {\cal D}\big\}\hfill&=&2{\cal H},\hfill
&\big\{{\cal H},{\cal K}\big\}\hfill&=&{\cal D},\hfill
&\big\{{\cal D},{\cal K}\big\}\hfill&=&2{\cal K},\hfill&
\\[8pt]
\big\{{\cal R},{\cal D}\big\}\hfill&=&0,\hfill
&\big\{{\cal R},{\cal K}\big\}\hfill&=&0,\hfill
&\big\{{\cal D},{\cal G}_i\big\}\hfill&=&{\cal G}_i,\hfill
\\[8pt]
\big\{{\cal D},{\cal P}_i\big\}\hfill&=&-{\cal P}_i,\hfill
&\big\{{\cal K},{\cal G}_i\big\}\hfill&=&0,\hfill
&\big\{{\cal K},{\cal P}_i\big\}\hfill&=&{\cal G}_i.\hfill
\end{array}
\label{Schrrel}
\end{equation}
In particular, $\cD$ and $\cK$ span, with
the energy, $\cH$, an $\O(2,1)$ subgroup.

Additional symmetries play an important r\^ole
\cite{JPDT}.
Deriving the expansion generator ${\cal K}$ in 
(\ref{SchrGen}) twice w.r.t. time shows in fact that
$$
\left(\frac{m}{2}\int r^2|\Psi|^2 d^2x\right)''
$$
is twice the Hamiltonian, and is hence time-independent. 
It follows
that for fields that make $|\Psi|^2$ time-independent,
in particular for static fields, the energy vanishes. 
Therefore, when $\Lambda$ takes the specific ``self-dual''
value
(\ref{specLambda}) the solution is necessarily self-dual
by eqn (\ref{NRBog}) \footnote{
In the
Abelian Higgs model the analogous theorem 
is rather difficult to prove \cite{Taubes}.}.

We mention for the record that
applying any symmetry transformation to a solution of the field equations 
yields another solution. For example, a boost or an expansion applied
to the static solution $\Psi_0(\vec{X})$ of Jackiw and Pi
produces time-dependent solutions, 
\begin{equation}
\Psi(T,\vec{X})=\frac{1}{1-kT}\exp\left\{-\frac{i}{2}\Big[
2\vec{X}\cdot\vec{b}+T\vec{b}^2+k\frac{(\vec{X}+\vec{b}T)^2}{1-kT}
\Big]\right\}\,\Psi_0(\frac{\vec{X}+\vec{b}T}{1-kT}).
\label{expansImp}
\end{equation}

%%%%%%%%%%%%%%%%%%%%%%%%%%%%%%%%%%%%%%%%%%%%%%%%%%%
\subsection{Symmetries in the non-relativistic Kaluza-Klein-type framework}\label{Bargmann}
%%%%%%%%%%%%%%%%%

How do the extra symmetries come about~?
Can one derive the
energy-momentum tensor (\ref{Tmunu}),
 together with its strange
properties (\ref{NRsymEM}), in a systematic way~?
This  is conveniently achieved in the ``non-relativistic
Kaluza-Klein'' framework.
The clue is that non-relativistic spacetime can be 
obtained from a $(3+1)$ dimensional relativistic spacetime,
$M$, endowed with a Lorentz-signature metric $g_{\mu\nu}$
and  a covariantly constant, lightlike vector $\xi^\mu$.
(Such a manifold, called a ``Bargmann space'',
is in fact a gravitational pp wave \cite{DGH}. These
spaces  can
provide exact string vacua \cite{string}).  
Then non-relativistic spacetime is the factor space of $M$,
obtained by factoring out the integral curves of $\xi^\mu$.
 
 When $M$ is the Minkowski space, in particular,
the metric can be written using light-cone coordinates 
$t$ and $s$ as
\begin{equation}
ds^2=d\vx{}^2+2dtds.
\label{Minkowski}
\end{equation}
More generally, we can have
\begin{equation}
ds^2=g_{ij}dx^idx^j+2dtd(s+A_idx^i)-2Udt^2,
\label{genBmetric}
\end{equation}
where $g_{ij}$ is some spatial metric
and $A_i$ and $U$ are a vector and a scalar potential,
respectively. 

The coordinate $\vx$ can be viewed as position, $t$ as
non-relativistic time, and $s$ as an ``internal,
Kaluza-Klein-type coordinate'', directed
along the ``vertical'' vector
$\xi^\mu=\p_s$. Quotienting $M$ by the integral curves
of $\xi^\mu$ amounts, intuitively, to ``forgetting'' $s$.

It is now easy to check that \textit{the projection of 
the null-geodesics} of 
$M$, endowed with the metric 
\begin{equation}
ds^2=g_{ij}dx^idx^j+2dtd(s+A_idx^i)-2Udt^2,
\label{emBmetric}
\end{equation}
 satisfy
the usual equations of motion of a non-relativistic
particle in a (static) ``electromagnetic'' field 
\begin{equation}
\vB=\rot\vA,
\qquad
\vE=-\grad U.
\end{equation}
With one strange detail, though~: the coupling constant is
not the electric charge, $e$, but the mass, $m$.

For $\vA=0$, in particular, we recover, as noticed by Eisenhart in 1929 \cite{Eisen}, Newton's equations. 

Null geodesics are conformally invariant and their
projections are hence invariant w.r.t. $\xi$-pre\-serving 
conformal transformations which are, hence,
symmetries of the projected system. 

In  Minkowski space (\ref{Minkowski}), in particular, the (infinitesimal)
conformal transformations span the conformal algebra
$\o(4,2)$; those which preserve the lightlike vector 
$\xi^\mu=\p_s$
are precisely the generators of the (centrally extended)
\textit{planar Schr\"odinger group},
centrally extended with the mass (the standard central
extension of the Galilei group).

This ``non-relativistic Kaluza-Klein'' framework
has been useful to study the Schr\"odinger symmetry of
classical systems, and can also adapted to CS field theory \cite{DHP1}.
Let us choose indeed, on $M$, a four-vector potential
$a_\mu$   with field strength $f_{\mu\nu}$ and let
$j_\mu$ be a four-current.

$\bullet$  Let us posit the relation
\begin{equation}
\kappa f_{\mu\nu}=e\sqrt{-g}\epsilon_{\mu\nu\rho\sigma}
\xi^\rho j^\sigma.
\label{4dFCI}
\end{equation}
Then $f_{\mu\nu}$ is the lift
from space-time with coordinates $\vx$ and $t$ of a
closed two-form $F_{\mu\nu}$. 
$a_\mu$ can be chosen therefore as the pull-back of a 3-potential
$A_\alpha=(A_t,\vA)$. The four-current $j^\mu$
projects in turn onto a 3-current $J^\alpha=(\varrho,\vJ)$. 
Then (\ref{4dFCI}) is readily seen to project precisely to
the Chern-Simons equations (\ref{Gauss})-(\ref{FCI}). 

 $\bullet$  Similarly, let $\psi$ denote a scalar field on $M$ and let
us posit the (massless) non-linear Klein-Gordon wave equation
\begin{equation}
\left[D_\mu D^\mu-\frac{R}{6}+\lambda(\psi^*\psi)\right]\psi=0,
\label{NLKG}
\end{equation}
where $D_\mu=\nabla_\mu-iea_\mu$ is the metric and
gauge covariant derivative on $M$ and we have also added,
for the sake of generality, a term which involves the
scalar curvature, $R$ of $M$. Requiring that the scalar field
be equivariant,
\begin{equation}
\xi^\mu D_\mu\psi=im\psi,
\label{equivar}
\end{equation}
$\Psi=e^{ims}\psi$ will be a function of $\vx$ and $t$ alone, and
(\ref{NLKG}) becomes, for the Minkowski
metric (\ref{Minkowski}), the gauged non-linear
Schr\"odinger equation (\ref{NLS}). 

$\bullet$  The systems (\ref{4dFCI}) 
and (\ref{NLKG}) are coupled through 
\begin{equation}
j^\mu=\frac{1}{2mi}\left[\psi^*(D^\mu\psi)-\psi(D^\mu\psi)^*\right],
\label{4dcur}
\end{equation}
that projects to the relation (\ref{NRcurrent}).

Eqns. (\ref{4dFCI})-(\ref{NLKG})-(\ref{equivar})-(\ref{4dcur}) form a self-consistent
system  allowing us to lift our non-relativistic
coupled scalar field-Chern-Simons system to the relativistic spacetime $M$.
It can be now shown \cite{DHP1}
that the latter is invariant
w.r.t. any conformal transformation of the metric of $M$
that preserves the ``vertical'' vector $\xi^\mu$.
Thus, we have just established the Schr\"odinger
invariance of the non-relativistic Chern-Simons + scalar field 
system.

The theory on $M$ is relativistic and admits, therefore,
a conserved, traceless and symmetric energy-momentum tensor
$\theta_{\mu\nu}$. In the present case, the canonical procedure
yields
\begin{equation}
\begin{array}{lll}
3m\theta_{\mu\nu}&=&
(D_\mu\psi)^*D_\nu\psi+D_\mu\psi(D_\nu\psi)^*
\\[12pt]
&&-\frac{1}{2}\left(
\psi^*D_\mu D_\nu\psi+\psi(D_\mu D_\nu)^*\right)
\\[12pt]
&&+\frac{1}{2}|\psi|^2\left(R_{\mu\nu}-\frac{R}{6}g_{\mu\nu}\right)
-\frac{1}{2}g_{\mu\nu}(D^\rho\psi)^*D_\rho\psi
-\frac{\lambda}{4}g_{\mu\nu}|\psi|^4.
\end{array}
\end{equation}
It is now easy to prove that
\begin{equation}
\begin{array}{ll}
T^{00}=-\theta^0_{\ 0},
&
T^{i0}=-\theta^i_{\ 0}-\frac{1}{6m}\p_i\p_t\varrho,
\\[10pt]
T^{0j}=\theta^0_{\ j}
&
T^{ij}=\theta^i_{\ j}+\frac{1}{3m}
\left(\delta^i_j\Delta-\p^i\p_j\right)\varrho,
\end{array}
\end{equation}
where $\Delta$ is the spatial Laplace operator. 
These formulae allow us
to infer all those properties of $T^{\alpha\beta}$ listed 
above. 

In Ref. \cite{DHP1}  a version of
Noether's theorem was proved. It says that, for any $\xi$-preserving
conformal vectorfield $(X^\mu)$ on Bargmann space, the quantity
\begin{eqnarray}
Q_X=\int_{\Sigma_t}
\vartheta_{\mu\nu}X^\mu\xi^\nu\,\sqrt{\gamma}\,d^2\vec{x},
\label{consQ}
\end{eqnarray}
is a constant of the motion. (Here 
$\gamma$ is the determinant
of the metric $g_{ij}$ induced by $g_{\mu\nu}$
on  `transverse space' $t={\rm const.}$.)
The charge (\ref{consQ}) is conveniently calculated using
\begin{eqnarray}
\vartheta_{\mu\nu}\xi^\nu
=\frac{1}{2i}\left[
\psi^*\,(D_\mu\psi)-\psi\,(D_\mu\psi)^*\right]
-\frac{1}{6}\,\xi_\mu\left(
\frac{R}{6}|\psi|^2
+(D^\nu\psi)^*\,D_\nu\psi
+\frac{\lambda}{2}\,|\psi|^4
\right).
\end{eqnarray}

It is worth mentionning that chosing the ``vertical'' vector
$\xi^\mu$ spacelike would provide us with a relativistic
theory ``downstairs''. 

It is interesting to note that our proof used the field equations.
Is it possible to extend it to the variational principle~?
On $M$ we could use in fact the $4$d 
``Chern-Simons type'' expression
\begin{equation}
\frac{\kappa}{2}\epsilon^{\mu\nu\rho\sigma}
\xi_\mu a_\nu f_{\rho\sigma}.
\label{4dCS}
\end{equation}
Curiously, this correctly reproduces the
{\it relativistic} Chern-Simons equations (\ref{relCSEL2}) if $\xi^\mu$
is spacelike, but fails when it is lightlike,
$\xi_\mu\xi^\mu=0$  \cite{HH} --- which is 
precisely the {\it non-relativistic} case we study here.

%%%%%%%%%%%%%%%%%%%%%%%%%%%%%%%%%%%
\subsection{Time-dependent vortices in an
external electromagnetic field}\label{t-dVort}
%%%%%%%%%%%%%%%%%%%%%%%%%%%%%%%%%%%

The  static, non-relativistic Chern-Simons solitons studied
above can be
generalized to yield time-dependent vortex solutions
in a constant external magnetic field
${\cal B}$ \cite{EHIt-d,JPt-d}. 
Putting $\omega={\cal B}/2$, the equation to be solved is
\footnote{We use here units where $e=m=1$.}
\begin{equation}
i\big({D_\omega}\big)_t\Psi_\omega=\left\{
-\frac{1}{2}{\vec{D}}_\omega^2
-\Lambda\,\Psi_\omega^*\Psi_\omega
\right\}\Psi_\omega.
\label{efNLS}
\end{equation}
Here the modified covariant derivative means
\begin{equation}
\big({D_\omega}\big)_\alpha
=
\partial_\alpha-i({A_\omega})_\alpha-i{\cal A}_\alpha
\label{Bcovder}
\end{equation}
($\alpha=0,1,2$), where ${\cal A}_\alpha$ is a vector potential for the constant magnetic field, chosen to be
$$
{\cal A}_0=0,
\qquad{\cal A}_i=
\2\epsilon_{ij}x^j{\cal B}\equiv\omega\epsilon_{ij}x^j
$$
($i,j=1,2$).
$(A_\omega)_\alpha$ is the ``statistical'' vector potential 
of Chern-Simons electromagnetism, whose field 
strength is required to satisfy the field-current identities
\begin{eqnarray}
B_\omega&\equiv&
\epsilon^{ij}\partial_iA_\omega^j
=-\frac{1}{\kappa}\varrho_\omega
\\[6pt]
E_\omega^i&\equiv&
-\partial_iA_\omega^0-\partial_tA_\omega^i
=\frac{1}{\kappa}\epsilon^{ij}J_\omega^j
\label{FCIt-d}
\end{eqnarray}
with
$\varrho_\omega=\Psi_\omega^*\Psi_\omega$
and
$\vec{J}_\omega=
({1/2i})[\Psi^*\vec{D}_\omega
\Psi_\omega-\Psi_\omega(\vec{D}_\omega\Psi_\omega)^*]
$.

These equations can be solved \cite{EHIt-d,JPt-d} applying a
coordinate transformation to a solution, $\Psi$ and $A_\alpha$, of the ``free'' problem
with $\omega=0$, according to
\footnote{The formulae in  \cite{JaPi1} also involve the factor
$\exp\left\{i\frac{{\cal N}}{2\pi\kappa}\omega t\right\}$,
as a result of gauge fixing.}
\begin{eqnarray}
\Psi_\omega(t,\vec{x})&=&\frac{1}{\cos\omega t}\,
\exp\left\{-i\omega\frac{r^2}{2}\tan{\omega t}\right\}\,
%\exp\left\{i\frac{{\cal N}}{2\pi\kappa}\omega t\right\}\,
\Psi(\vec{X},T),\label{psiexport}
\\[12pt]
(A_\omega)_\alpha&=&A_\beta\frac{\partial X^\beta}{\partial x^\alpha},
%-\partial_\alpha\Big(\frac{\omega}{2\pi\kappa}{\cal N}t\Big),
\label{Aexport}
\end{eqnarray}
with
\begin{equation}
T=\frac{\tan\omega t}{\omega},
\qquad
\vec{X}=\frac{1}{\cos\omega t}R(\omega t)\,\vec{x}.
\label{TtXx}
\end{equation}
where 
%${\cal N}=\int|\Psi|^2\,d^2\vec{x}$ is the vortex number  and 
$R(\theta)$ is the matrix of a planar rotation with angle 
$\theta$.

A similar construction works in a harmonic background 
\cite{JPt-d}.

Now we explain the above results in  our ``Kaluza-Klein-type''
framework introduced in the previous Section.
Let us indeed consider coupled system  
(\ref{4dFCI})-(\ref{NLKG})-(\ref{equivar})-(\ref{4dcur})
on a  general ``Bargmann'' metric (\ref{emBmetric}).
Easy calculation shows that, 
after reduction,  the covariant derivative 
\begin{equation}
D_\alpha=\nabla_\alpha-iea_\alpha
\label{gmcovder}
\end{equation}
(where $\nabla_\alpha$ is the gauge-covariant
derivative) becomes precisely
${(D_\omega)}_\alpha$ in (\ref{Bcovder}), with, perhaps,
a nontrivial $A_0$.
The equation of motion is therefore an
 generalization of (\ref{efNLS}). 
 
$\bullet$ Let us consider, for example, the `` oscillator''
metric
\begin{equation}
d\vec{x}_{osc}^2+2dt_{\rm osc}ds_{\rm osc}-\omega^2r_{\rm osc}^2dt_{\rm osc}^2,
\label{oscMet}
\end{equation}
where $\vec{x}_{\rm osc}\in\IR^2$, $r_{\rm osc}=|\vec{x}_{\rm osc}|$ 
and $\omega$ is a constant. Its null geodesics  
correspond in fact to a non-relativistic,
spinless particle in an oscillator background \cite{DGH}.
Requiring equivariance, (\ref{equivar}), the wave equation 
(\ref{NLKG}) reduces to 
\begin{equation}
i\partial_{t_{\rm osc}}\Psi_{\rm osc}=\left\{
-\frac{\vec{D}^2}{2}+\frac{\omega^2}{2}{r_{\rm osc}}^2
-\Lambda\,\Psi_{\rm osc}\Psi_{\rm osc}^*\right\}\Psi_{\rm osc}
\end{equation}
($\vec{D}=\vec{\partial}-i\vec{A}$, $\Lambda=\lambda/2$),
which describes Chern-Simons vortices in a harmonic-force background, studied in Ref. \cite{JPt-d}.

$\bullet$ Let us consider instead the ``magnetic'' metric
\begin{equation}
d\vec{x} {}^2+2dt\Big[ds+
\2\epsilon_{ij}{\cal B}{x}^jd{x}^i\Big],
\label{BMet}
\end{equation}
where $\vec{x}\in\IR^2$ and ${\cal B}$ 
is a constant, whose null geodesics
describe a charged particle in a uniform magnetic 
field in the plane \cite{DGH}. 
Imposing equivariance, Eq.~(\ref{NLKG}) 
reduces  
to Eq.~(\ref{efNLS}) with $\Lambda=\lambda/2$ and the
covariant derivative $D_\omega$  in Eq.~(\ref{Bcovder}).

Returning to the general theory, let $\varphi$ denote a conformal 
Bargmann diffeomorphism between \textit{two} 
Bargmann spaces, i.e. let $\varphi~:~(M,g,\xi)\to(M',g',\xi')$ be
such that 
\begin{equation}
\varphi^\star g'=\Omega^2g
\qquad
\xi'=\varphi_\star\xi.
\label{Bconftransf}
\end{equation}
Such a mapping projects to a diffeomorphism of the quotients,
$Q$ and $Q'$ we denote by $\Phi$.
Then the same proof as in Ref. \cite{DHP1} allows one to show that if 
$(a'_\mu,\psi')$ is a solution of the field equations on 
$M'$, then 
\begin{equation}
a_\mu=(\varphi^\star a')_\mu
\qquad
\psi=\Omega\,\varphi^\star\psi'
\label{apsiexp}
\end{equation} 
is a solution of the analogous equations on $M$. Locally 
$$
\varphi(t,\vec{x},s)=(t',\vec{x}',s')\quad\hbox{with}\quad
(t',\vec{x}')=\Phi(t,\vec{x}),
\quad
s'=s+\Sigma(t,\vec{x}),
$$ 
so that
$\psi=\Omega\,\varphi^\star\psi'$ reduces to
\begin{equation}
\Psi(t,\vec{x})=\Omega(t)\,e^{i\Sigma(t,\vec{x})}\Psi'(t',\vec{x}'),
\qquad
A_\alpha=\Phi^\star A'_\alpha
\end{equation}
($\alpha=0,1,2$).
Note that $\varphi$ takes a $\xi$-preserving conformal 
transformation of $(M,g,\xi)$ into a $\xi'$-preserving conformal
transformation of $(M',g',\xi')$.
Conformally related Bargmann spaces
have therefore isomorphic symmetry groups.

The conserved quantities can be related by comparing 
the expressions in (\ref{consQ}). 
Using the transformation properties of the scalar curvature,
 short calculation shows that the conserved quantities
associated to $X=(X^\mu)$ on $(M,g,\xi)$ and to 
$X'=\varphi_\star X$ on 
$(M',g',\xi')$ coincide,
\begin{equation}
Q_X=\varphi^\star Q'_{X'}.
\label{Qexport}
\end{equation}
The labels of the generators
are, however, different (see the examples below).

$\bullet$ As a first application, we note the 
 the lift to Bargmann space of Niederer's mapping
\cite{NiedererOsc}
\begin{eqnarray}
&\varphi(t_{\rm osc},\vec{x}_{\rm osc},s_{\rm osc})
=(T,\vec{X},S),&\nonumber
\\[10pt]
&T=\displaystyle\frac{\tan\omega\,t_{\rm osc}}{\omega},
\qquad
\vec{X}= \displaystyle\frac{\vec{x}_{\rm osc}}{\cos\omega t_{\rm osc}},
\qquad
&S=s_{\rm osc}-\displaystyle\frac{\omega r_{\rm osc}^2}{2}\tan\omega t_{\rm osc}
\label{oscfree}
\end{eqnarray}
carries the oscillator metric (\ref{oscMet}) 
Bargmann-conformally 
($\varphi_\star\partial_{s_{\rm osc}}=\partial_S$)
into the free form (\ref{Minkowski}),
with conformal factor 
$\Omega(t_{\rm osc})=|\cos\omega t_{\rm osc}|^{-1}$. 
 A solution in the harmonic background can be obtained by 
Eq.~(\ref{apsiexp}).
 
A subtlety arises, though: the mapping (\ref{oscfree}) is many-to-one~:  it maps each `open strip' 
\begin{equation}
I_j=\big\{
(\vec{x}_{\rm osc},t_{\rm osc},s_{\rm osc})\,\big|
\,(j-\2)\pi<\omega t_{\rm osc}<(j+\2)\pi
\big\},
\qquad
\,j=0,\pm1,\ldots
\end{equation}
corresponding to a \textit{half oscillator-period}, onto 
full Minkowski space. 
Application of (\ref{apsiexp}) with $\Psi$ an `empty-space' solution
yields, in each $I_j$, a solution, $\Psi^{(j)}_{\rm osc}$. However,
at the contact points 
$t_j\equiv(j+1/2)(\pi/\omega)$, these fields may not match.
For example, for the `empty-space' solution obtained by an expansion, 
Eq.~(\ref{expansImp}) with $\vec{b}=0,\,k\neq0$, 
\begin{equation}
\lim_{t_{\rm osc}\to t_j-0}\Psi^{(j)}_{\rm osc}=
(-1)^{j+1}\frac{\omega}{ k}
e^{-i\frac{\omega^2}{2k}r_{\rm osc}^2}\Psi_0(-\frac{\omega}{ k}\vec{x})=
-\lim_{t_{\rm osc}\to t_j+0}\Psi^{(j+1)}_{\rm osc}.
\end{equation}
The lef-and right limits differ hence by a sign.
The continuity of the wave functions is restored including the `Maslov' phase correction \cite{Maslov}~:
\begin{equation}
\begin{array}{lll}
\Psi_{\rm osc}(t_{\rm osc},\vec{x}_{\rm osc})&=&
(-1)^{j}\,\displaystyle\frac{1}{\cos\omega t_{\rm osc}}\,
\exp\left\{-\frac{i\omega}{2}r_{\rm osc}^2\tan{\omega t_{\rm osc}}\right\}\,
\Psi(T,\vec{X})
\\[18pt]
(A_{\rm osc})_0(t_{\rm osc},\vec{x}_{\rm osc})
&=& \displaystyle\frac{1}{\cos^2\omega t_{\rm osc}}
\big[
A_0(T,\vec{X})-\omega\sin\omega t_{\rm osc}\;
\vec{x}_{\rm osc}\cdot\vec{A}(T,\vec{X})
\big],
\\[18pt]
\vec{A}_{\rm osc}(t_{\rm osc},\vec{x}_{\rm osc})
&=& \displaystyle\frac{1}{\cos\omega t_{\rm osc}}\,
\vec{A}(T,\vec{X}),
\end{array}
\label{MaslovExport}
\end{equation}
 
Eq.~(\ref{MaslovExport}) 
extends the result in \cite{JPt-d}, which are only valid for 
$|t_{\rm osc}|<\pi/2\omega$,
to any $t_{\rm osc}$ (\footnote{For the static solution in \cite{JaPi1} 
or for that obtained from it by 
a boost, $\lim_{t_{\rm osc}\to t_j}\Psi^{(j)}_{\rm osc}=0$,
and the inclusion of the correction factor is not mandatory.
}).

Since the oscillator metric (\ref{oscMet}) is 
Bargmann-conformally related to
Minkowski space, Chern-Simons theory in the oscillator background 
has again a Schr\"odinger symmetry -- but with ``distorted''
generators. The latter are in fact
\begin{eqnarray}
J_{\rm osc}=\cJ,
\qquad
H_{\rm osc}={\cH}+\omega^2\cK,
\qquad
N_{\rm osc}=\cN
\end{eqnarray}
completed by
\begin{eqnarray}
(C_{\rm osc})_\pm=
\left(\cH-\omega^2\cK\pm 2i\omega\,\cD\right),
\qquad%\\[8pt]
(\vec{P}_{\rm osc})_\pm=
\left(\vec{\cP}\pm i\omega\,\vec{\cG}\right).
\end{eqnarray}
Let us observe in particular
that the oscillator-Hamiltonian, $H_{\rm osc}$, 
is  a combination of the ``empty-space'' !$\omega=0$)
Hamiltonian and  expansion, etc. 
%Eq.~(20) adds $(C_{\rm osc})_\pm$ and $(\vec{P}_{\rm osc})_\pm$
%to the $J_{\rm osc}$ and $H_{\rm osc}$. 

$\bullet$ Turning to the magnetic case, let us
observe that the ``magnetic'' metric (\ref{BMet}) is 
readily transformed into an oscillator metric (\ref{oscMet}),
namely by the mapping 
$
\varphi(t,\vec{x},s)=(t_{\rm osc},\vec{x}_{\rm osc},s_{\rm osc}),
$
\begin{equation}
t_{\rm osc}=t,
\qquad
x_{\rm osc}^i=x^i\cos\omega t+\epsilon^i_jx^j\sin\omega t,
\qquad
s_{\rm osc}=s
\label{tdrot}
\end{equation}
[which amounts to switching 
to a rotating frame with angular velocity $\omega={\cal B}/2$]. 
The vertical vectors
$\partial_{s_{\rm osc}}$ and $\partial_s$ are permuted.

Composing the two steps, we see that the time-dependent rotation (\ref{tdrot}), 
followed by the transformation (\ref{oscfree}),
[which projects to the coordinate transformation (\ref{TtXx})], 
carries conformally the 
constant-${\cal B}$ metric (\ref{BMet}) into the free
($\omega=0$)-metric. 
It carries therefore 
the \lq empty' space so\-lution $e^{is}\Psi$ with $\Psi$ as
in (\ref{expansImp}) into that in
a uni\-form mag\-netic field back\-ground
according to Eq.~(\ref{apsiexp}). Taking into account the equivariance,
we get the formul{\ae} of \cite{EHIt-d}, multiplied with the Maslov factor $(-1)^j$. 

Our framework also allows to
\lq export' the Schr\"odinger symmetry to
non-relativistic Chern-Simons theory in
the constant magnetic field background. The (rather 
complicated) generators \cite{Hotta} can be
 obtained using Eq.~(\ref{Qexport}). 
For example, time-translation $t\to t+\tau$ in 
the ${\cal B}$-background amounts to a time translation for the 
oscillator with parameter $\tau$ plus a rotation with angle $\omega\tau$.
Hence
$$
H_{\cal B}=H_{\rm osc}-\omega\cJ=\cH+\omega^2\cK-\omega\cJ.
$$

Similarly, a space translation for ${\cal B}$ amounts, in `empty' space,
to a space translations and a rotated boost~: 
$P_B^i=\cP^i+\omega\,\epsilon^{ij}\cG^j$, etc.

All our preceding results apply to any Bargmann space which can
be Bargmann-conformally mapped into Minkowski space.
Now we describe all these 
`Schr\"odinger-conformally flat' spaces.
In $D=n+2>3$ dimensions, conformal
flatness is guaranteed by the
vanishing of the conformal Weyl tensor $C^{\mu\nu}_{\ \ \rho\sigma}$.
%A tedious calculation \cite{DHP2} yields
%\begin{equation}
%C^{\mu\nu}_{\ \ \rho\sigma}=
%R^{\mu\nu}_{\ \ \rho\sigma}
%-\frac{4}{D-2}\,
%\varrho\,\delta^{[\mu}_{\ [\rho}\,\xi^{\nu]}\xi^{ }_{\sigma]}.
%\end{equation}
 Skipping technical details,
we state that Schr\"odinger-conformal flatness requires \cite{DHP2}
\begin{eqnarray}
&{\cal A}_i=\2\epsilon_{ij}{\cal B}(t)x^j+a_i,
\qquad
\vec{\nabla}\times\vec{a}=0,
\qquad\partial_t\vec{a}=0,
\\[8pt]
&U(t,\vec{x})=\2 C(t)r^2+\vec{F}(t)\cdot\vec{x}+K(t).
\label{confflat}
\end{eqnarray}

The  metric (\ref{genBmetric})-(\ref{confflat}) describes a
uniform magnetic field ${\cal  B}(t)$, 
an attractive [$C(t)=\omega^2(t)$] or repulsive 
[$C(t)=-\omega^2(t)$]
isotropic oscillator and a uniform force field $\vec{F}(t)$ in the plane, all of
which may depend on time. It also includes a curlfree 
vector potential~$\vec{a}(\vec{x})$ that can be gauged away if the 
transverse space is simply connected: $a_i=\partial_if$ and the coordinate transformation $(t,\vec{x},s)\to(t,\vec{x},s+f)$
results in the `gauge' transformation 
\begin{equation}
{\cal{A}}_i\to{\cal{A}}_i-\partial_if=
-\2{\cal B}\,\epsilon_{ij}x^j.
\end{equation} 
If, however, space is not simply connected, we can 
also include an external Aharonov-Bohm-type vector potential. 

Being conformally related, all these metrics share the symmetries of flat 
Bargmann space: for example, if the transverse space is $\IR^2$ we get the 
full Schr\"odinger symmetry; for $\IR^2\setminus\{0\}$ the symmetry is reduced rather to
${\rm o}(2)\times{\rm o}(2,1)\times\IR$,
as found for a magnetic vortex \cite{magvort}.

The case of a constant electric field 
is quite amusing. Its metric,
$d\vec{x}^2+2dtds-2\vec{F}\cdot\vec{x}dt^2$, 
can be brought to the free 
form by switching to an accelerated coordinate system,
\begin{equation}
\vec{X}=\vec{x}+\2\vec{F}\,t^2,
\quad
T=t,
\quad
S=s-\vec{F}\cdot\vec{x}\,t-\smallover1/6\vec{F}^2t^3.
\label{cEfield} 
\end{equation}
This example 
also shows that the action of the Schr\"odinger group --- e.g. a rotation 
--- looks quite differently in the inertial and in the moving frames. 

In conclusion, our `non-relativistic Kaluza-Klein' approach provides 
a unified view on the various known constructions  
and explains the common origin of their symmetries.
%All such conformally flat Bargmann spaces were determined.

%%%%%%%%%%%%%%%%%%%%%%%%%%%%
%%%%%%%%%%%%%%%%%%%%%%%%%%%%%%%%%%%
\section{Non-relativistic Maxwell-Chern-Simons 
Vortices}\label{background}
%%%%%%%%%%%%%%%%%%%%%%%%%%

Generalizing previous work \cite{BaHa,DoIe},
 Manton \cite{Mantonmodel} proposed a modified version
of the Landau-Ginzburg model for describing
Type II superconductivity.
His Lagrange  density is a subtle mixture blended from
 the usual Landau-Ginzburg expression, augmented with
the Chern-Simons term:
\begin{eqnarray}
{\cal L}=
&-\frac{1}{2}B^2+
\gamma\frac{i}{2}\big(\phi^*D_t\phi-\phi(D_t\phi)^*\big)
-\frac{1}{2}\big|\vec{D}\phi\big|^2
-\frac{\lambda}{8}\big(1-|\phi|^2\big)^2\nonumber
\\[8pt]
&+\mu\big(Ba_t+E_2a_1-E_1a_2\big)
-\gamma a_t-\vec{a}\cdot\vec{J}^{T},\label{ManLag}
\end{eqnarray}
where $\mu$, $\gamma>0$, $\lambda>0$ are constants,
$D_t\phi=\partial_t\phi-ia_t\phi$ and
$D_i\phi=\partial_i\phi-ia_i\phi$ are the partial derivates,
$B=\partial_1a_2-\partial_2a_1$ is the magnetic field
and $\vec{E}=\vec\nabla a_t-\partial_t\vec{a}$
is the electric field.

This Lagrangian  has the usual symmetry-breaking quartic
potential, but differs from the standard expression in that
\begin{enumerate}
\item
it is linear in $D_{t}\phi$
but quadratic in $D_{i}\phi$;
\item  
the Maxwellian electric term $\vec{E}^2$ is missing; 
\item includes the ``weird'' terms
$-\gamma a_{t}$ and
$-\vec{a}\cdot\vec{J}^{T}$, where $\vec{J}^{T}$
is the (constant) transport current. 
\end{enumerate}

Properties (1) and (2) stem from the requirement of Galilean
rather than Lorentz invariance. 
The term $-\gamma a_{t}$ results in modifying
the Gauss law (eqn. (\ref{ManGauss}) below);
the term $-\vec{a}\cdot\vec{J}^{T}$
is then needed in order to restore the Galilean invariance.
To be so, the transport current has to transform as
 $\vec{J}^{T}\to\vec{J}^{T}+\gamma\vec{v}$
under a Galilei boost \cite{Mantonmodel}.

The field equations derived from (\ref{ManLag}) read 
\begin{eqnarray}
&i\gamma D_t\phi=
-\frac{1}{2}\vec{D}^2\phi
-\frac{\lambda}{4}\big(1-|\phi|^2\big)\phi,
\label{ManNLS}
\\[8pt]
&\epsilon_{ij}\partial_{j}B=
J_{i}-J^{T}_{i}+2\mu\,\epsilon_{ij}\,E_j,
\label{ManFCI}
\\[8pt]
&2\mu B=\gamma\big(1-|\phi|^2\big),
\label{ManGauss}
\end{eqnarray}
where the (super)current 
is
$
\vec{J}=({1/2i})\big(\phi^*\vec{D}\phi-\phi(\vec{D}\phi)^*\big).
$

$\bullet$ The matter field satisfies hence a
gauged, planar non-linear Schr\"odinger equation.

$\bullet$ The second equation is Amp\`ere's
law without the displacement current, as usual in
the ``magnetic-type'' Galilean electricity \cite{LBLL}.

$\bullet$ The last equation (which replaces the Gauss law of Maxwellian dynamics)
is the (modified) ``Field-Current Identity''.

Manton \cite{Mantonmodel} observed that when
$\vec{J}^T=0$,
 $\lambda=1$ and $\mu=\gamma$, these same solutions
yield magnetic vortices 
with $a_{t}=0$, also in the Chern-Simons-modified model. 
Below, we  generalize Manton's results to construct
solutions with a non-vanishing electric field.
 
Before searching for solutions, let us  discuss  the finite-energy 
conditions.  In the frame where $\vec{J}^{T}=0$,
the energy associated to the Lagrangian (\ref{ManLag}) is \cite{HHY} 
\begin{equation}
H=\int\Big\{
\2\big\vert\vec{D}\phi\big\vert^{2}+\2B^{2}
+U(\phi)\Big\}
\,d^{2}\vec{x},
\qquad
U(\phi)=\frac{\lambda}{8}\big(1-\vert\phi\vert^{2}\big)^{2}.
\label{Manenergy1}
\end{equation}
Eliminating the magnetic term $B^2/2$ 
using the Gauss law (\ref{ManGauss})
results in a mere shift of the coefficient of the non-linear 
term,
\begin{equation}
H=\int\Big\{
\2\big\vert\vec{D}\phi\big\vert^{2}
+
\frac{\Lambda}{8}\big(1-\vert\phi\vert^{2}\big)^{2}\Big\}
\,d^{2}\vec{x},
\label{Manenergy}
\qquad
\Lambda=\lambda+\frac{\gamma^{2}}{\mu^{2}}.
\end{equation}

Finite energy ``requires'', just like in the 
Landau-Ginzburg case,
\begin{equation}
\vec{D}\phi\to0
\qquad\hbox{and}\qquad
\vert\phi\vert^2\to1,
\label{topMCS}
\end{equation}

%%%%
By eqn. (\ref{topMCS}) we get, hence, \textit{topological vortices}~:
 the asymptotic values of scalar field provide us with a
mapping from
the circle at infinity ${S}$
into the vacuum manifold $|\phi|^2=1$ which is again a circle,
\begin{equation}
\psi\Big|_{\infty}: S\to S^1.
\end{equation}
\begin{figure}
\begin{center}
\includegraphics[scale=0.8]{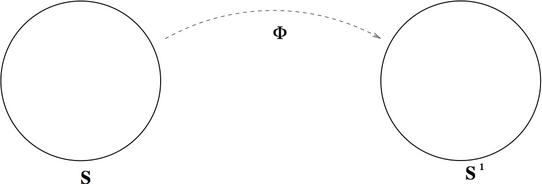}
\end{center}\vspace{-5mm}
\caption{\it  The asymptotic values of the scalar field provide us
with a mapping of the ``circle at infinity'' into the
unit cercle $|\psi|=1$. The winding number of the mapping
is the topological charge, related to quantized magnetic flux.}
\end{figure} 
%%%%
The first of the  equations in (\ref{topMCS}) 
implies that the angular component 
of vector potential behaves asymptotically as $n/r$.
The integer $n$ here is also the
winding number of the mapping defined by the asymptotic values
of $\phi$ into the unit circle,
\begin{equation}
n=\frac{1}{2\pi}\oint_{S}\vec{a}\cdot d\vec{\ell}
=\frac{1}{2\pi}\int B\,d^2\vec{x}.
\end{equation}

The \textit{magnetic flux is therefore quantized} and is related to
the  particle number 
\begin{equation}
N\equiv\int\big(1-\vert\phi\vert^2\big)\,d^2\vec{x}
=\frac{2\mu}{\gamma}\int B\,d^2\vec{x}
=4\pi\big(\frac{\mu}{\gamma}\big)\,n
\end{equation}
by (\ref{ManGauss}).
$N$ is conserved since the supercurrent satisfies the 
continuity equation
$
\partial_t\varrho+\vec\nabla\cdot\vec{J}=0
$.

\goodbreak
%%%%%%%%%%%%%%%%%%%%%%%%%%%%%%%%%%%%%%%%%%%%%%%%%%
\subsection{Self-dual Maxwell-Chern-Simons vortices}
%%%%%%%%%%%%%%%%%%%%%%%%%%%%%%%%%%%%%%%%%%%%%%%%%

Conventional Landau-Ginzburg theory admits finite-energy, static, 
 purely magnetic vortex solutions.
For a specific value of the coupling constant, one can find solutions
by solving instead the first-order ``Bogomolny'' equations \cite{Abrik,NiOl,Bogo}, 
\begin{equation}
\begin{array}{clc}
(D_1+iD_2)\phi&=&0,
\\[6pt]
2B&=&1-\vert\phi\vert^2.
\end{array} \label{ManManSD}
\end{equation}

 In the frame where $\vec{J}^{T}=0$ (which 
can always be achieved by a Galilei boost),
the static Manton equations read 
\begin{equation}\begin{array}{ll}
&\gamma a_t\phi=
-\frac{1}{2}\vec{D}^2\phi
-\displaystyle\frac{\lambda}{4}\big(1-|\phi|^2\big)\phi,
\\[8pt]
&\vec\nabla\times B=\vec{J}+2\mu\vec\nabla\times a_t,
\\[8pt]
&2\mu B=\gamma\big(1-|\phi|^2\big).
\end{array}
\end{equation} 
Let us try  to solve these equations by the first-order Ansatz
\begin{equation}\begin{array}{l}
(D_{1}\pm iD_2)\phi=0,
\\[8pt]
2\mu B=\gamma\big(1-|\phi|^2\big).
\end{array}\label{ManSD}
\end{equation}
From the first of these relations we infer that
$$
\vec{D}^2=\mp i\big[D_1, D_2\big]=\mp B
\and
\vec{J}=\mp\2\vec\nabla\times\varrho,
$$
where $\varrho=|\phi|^2$.
 Inserting into the non-linear Schr\"odinger equation
we find that it is identically satisfied when 
$$
a_t=(\pm{1/4\mu}-{\lambda/4\gamma})(1-\varrho).
$$
Then from Amp\`ere's law we get that $\lambda$ has to be
\begin{equation}
\lambda=\pm2\frac{\gamma}{\mu}-\frac{\gamma^2}{\mu^2}.
\label{specManlambda}
\end{equation}
The scalar potential is thus
\begin{equation}
a_t=
\smallover1/{4\mu}\big(\mp1+\smallover{\gamma}/{\mu}\big)\,
\big(1-\varrho\big).
\label{Manat}
\end{equation}

The vector potential is expressed using the ``self-dual'' 
(SD) Ansatz (\ref{ManSD}) as
\begin{equation}
\vec{a}=\pm\2\vec\nabla\times\log\varrho+\vec\nabla\omega,
\label{Manvecpot}
\end{equation}
where $\omega$ is an arbitrary real function
chosen so that $\vec{a}$ is regular.
Inserting this into the Gauss law,  we end up with
the ``Liouville-type'' equation
$$
\bigtriangleup\log\varrho=\pm\alpha\big(\varrho-1\big),
\qquad
\alpha=\frac{\gamma}{\mu}.
$$

Now, if we want a ``confining'' (stable) and bounded-from-below scalar potential, $\lambda$ has to be positive. 
Then we see from eq. (\ref{specManlambda}) that for 
the upper sign this means $0<\alpha<2$,
 whereas for the lower
sign $-2<\alpha<0$.
In any of the two cases ($\alpha$ positive or negative), the 
coefficient of $(\varrho-1)$ in the r. h. s.
 is always positive: in the upper sign, it is
$\alpha$ with $\alpha>0$, in the lower sign,
it is $-\alpha$ with $\alpha<0$.
We consider henceforth 
\begin{equation}
\bigtriangleup\log\varrho=\vert\alpha\vert\big(\varrho-1\big);
\label{ManLioutype}
\end{equation}
the magnetic and electric fields can be obtained from
(\ref{Manvecpot}) and (\ref{Manat}).
Note that the electric field, $\vec{E}=\vec{\nabla}a_{t}$,
only vanishes for $\mu=\pm\gamma$, i.e., when $\lambda=1$,
which is Manton's case.

The self-duality equations (\ref{ManSD}) can also be obtained
by studying the energy, (\ref{Manenergy}). 
Using the identity 
$$
\big|\vec{D}\phi\big|^2=
\big|(D_1\pm iD_2)\phi\big|^2
\pm B|\phi|^{2}
\pm\vec{\nabla}\times\vec{J}
$$
and assuming that the fields vanish at infinity, the integral of
the current-term can be dropped, so that $H$ becomes 
\begin{equation}
\int\bigg\{
\frac{1}{2}\Big|(D_1\pm iD_2)\phi\Big|^2
+\Big[
\big(\mp\frac{\gamma}{4\mu}+
\frac{\Lambda}{8}\big)(1-|\phi|^2)^2\Big]
\bigg\}d^2\vec{x}
\pm\underbrace{\2\int B\, d^2\vec{x}}_{\pi n},
\label{ManEnBogodecomp}
\end{equation}
which shows that the energy is positive definite  when the square 
bracket vanishes, i.e., 
for the chosen potential with the 
special value (\ref{specManlambda}) of $\lambda$.
In this case, the energy admits a lower ``Bogomolny'' bound,
$H\geq\pi|n|$, with the equality only attained when the
SD equations hold.

Eqn. (\ref{ManLioutype}) is essentially  that of Bogomolny
in the Landau-Ginzburg theory \cite{Bogo}, to which it reduces
when $\vert\alpha\vert=1$. 
The proofs of  Weinberg \cite{Weinberg}, and of Taubes \cite{Taubes},
carry over literally to show,  for each $n$,
the \textit{existence of a $2n$-parameter family of solutions}.

Radial solutions can be studied numerically \cite{BaHa}; they 
 behave roughly as in the Bogomolny case. Write  $\phi=f(r)e^{in\theta}$ where $(r,\theta)$ are polar
coordinates in the plane.
Linearizing the Liouville-type eqn. (\ref{ManLioutype}),
we get for  the  deviation from the vacuum value, $\varphi=1-f$,
\begin{equation}
\varphi''+\frac{1}{r}\varphi'-\vert\alpha\vert\varphi=0,
\end{equation}
which is Bessel's equation of order zero. The
solution and its asymptotic behaviour are therefore
\begin{equation}
1-\varphi(r)\;\sim\;
\begin{array}{llll}
1-K_{0}(mr)
&\sim&
1-\displaystyle\frac{C}{\sqrt{r}}e^{-mr},
\qquad
&m=\sqrt{\vert\alpha\vert}.
\label{Manlarger}
\end{array}
\end{equation}
\begin{figure}
\begin{center}
\includegraphics[scale=0.62]{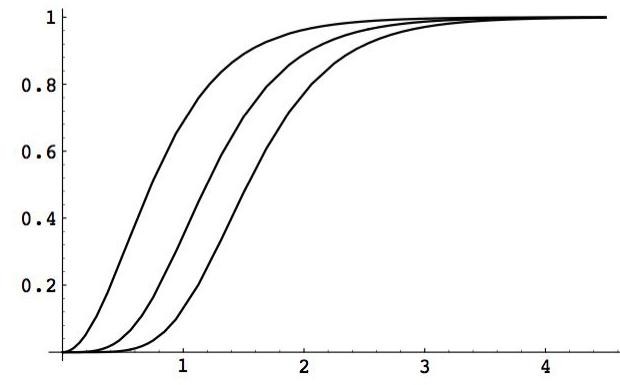}
\end{center}\vspace{-6mm}
\label{top}
\caption{\it The scalar field of the radially symmetric
Maxwell-Chern-Simons vortices with winding numbers
$n=1,2,3$.}
\end{figure}
It is, however, more convenient to study the first-order equations
instead of the Liouville-type eqn. (\ref{ManLioutype}).
For the radial Ansatz
\begin{equation}
a_r=0,
\qquad
a_\theta=a(r)
\end{equation}
the self-duality equations read indeed
\begin{equation}
f'=\pm\frac{n+a}{r}\,f,
\qquad
\frac{a'}{r}=\pm2(f^2-1).
\end{equation}
For small $r$ we get 
\begin{equation}
f(r)\sim \beta r^{|n|},
\qquad
a\sim\mp r^2
\end{equation}
where $\beta$ is some real parameter.
The large-$r$ behaviour (\ref{Manlarger}) of the scalar field
is confirmed, and for the magnetic and electric fields we get
\begin{eqnarray}
B&=&\frac{\alpha}{2}(1-f^2)\sim{\alpha}
\frac{D}{\sqrt{r}}
e^{-mr},
\\[6pt]
\vec{E}&=&-\smallover1/{4\mu}\big(\mp1+\alpha\big)\,
\vec\nabla f^2\sim
\frac{G}{\sqrt{r}}e^{-mr}.
\end{eqnarray}

Let us mention that the symmetries of the Manton model can
be studied along the lines indicated above. The clue is 
 to observe that putting
\begin{equation}
B^{ext}\equiv
\frac{\gamma}{2\mu},
\qquad
E^{ext}_{k}=-\frac{\epsilon_{kl}{J^{T}}_{l}}{2\mu},
\end{equation}
transforms the equations of motion 
(\ref{ManNLS}-\ref{ManFCI}-\ref{ManGauss}) into
\begin{equation}\begin{array}{llll}
&i\gamma D_t\phi&=
&-\frac{1}{2}\vec{D}^2\phi
-\frac{\lambda}{4}\big(1-|\phi|^2\big)\phi,
\\[8pt]
&\epsilon_{ij}\partial_{j}\widetilde{B}&=
&J_{i}+2\mu\,\epsilon_{ij}\,\widetilde{E}_j,
\\[8pt]
&2\mu\widetilde{B}&=
&-\gamma|\phi|^2,
\end{array}
\label{extMan}
\end{equation}
\vspace{-3mm}where \vspace{-3mm}
\begin{eqnarray}
\widetilde{B}=B-B^{ext}
\qquad
\widetilde{E}_{i}=E_{i}-E^{ext}_{i},
\\[4pt]
D_{\alpha}=\partial_{\alpha}-ia_{\alpha}, 
\qquad
a_{\alpha}=\tilde{A}_{\alpha}+A_{\alpha}^{ext}.
\end{eqnarray} 
These equations describe
a non-relativistic scalar field
with Maxwell-Chern-Simons dynamics with a symmetry-breaking quartic potential, put into a constant external
electromagnetic field. 
For details and for a discussion
of the Manton model to other similar ones  
 \cite{BaHa,DoIe},  the reader is referred to \cite{HHY}.

\goodbreak
%%%%%%%%%%%%%%%%%%%%%%%%%%%%%%%%%%%%%%%%%%%%%%%%%%%%%%%%%%%%%%
\subsection{Relativistic models and their non-relativistic limit}\label{relManton}
%%%%%%%%%%%%%%%%%%%%%%%%%%%%%%%%%%%%%%%%%%%%%%%%%%%%%%%%%%%%%%

In relativistic Maxwell-Chern-Simons theory
self-dual solutions only arise  when an auxiliary neutral field 
$N$ is added \cite{LLM}. Here we present a
  model of this type,  which
(i) is relativistic; (ii) can be made self-dual;
(iii) its non-relativistic limit is the Manton model
presented in this paper.
Let us consider in fact $(1+2)$-dimensional Minkowski space
with the metric  $(c^{2}/\gamma,-1,-1)$ where $\gamma>0$ is a constant.
Let us choose the Lagrangian
\begin{equation}
\L_{R}=-\smallover1/4F_{\mu\nu}F^{\mu\nu}
+
\smallover\mu/2\epsilon^{\mu\nu\rho}F_{\mu\nu}a_\rho
+
\big(D_\mu\psi\big)\big(D^\mu\psi\big)^*
+
a^{\mu}{J^T}_{\mu}
+
\smallover{\gamma}/{2c^{2}}\partial_\mu N\partial^\mu N
-V.
\end{equation}

Here $N$ is  an auxiliary neutral field, which we choose real. 
We have also included the term $a^{\mu}{J^T}_{\mu}$,
where the Lorentz vector ${J^T}_{\mu}$ represents the relativistic 
generalization of Manton's transport current. We choose ${J^T}_{\mu}$
to be time-like, $\I^{2}\equiv
\frac{\gamma}{ c^2}{J^T}_{\mu}{J^T}^{\ \mu}>0$.
 Our choice for the potential is
\begin{equation}
V=\frac{\beta}{2}\big(\vert\psi\vert^{2}-2\vert\mu\vert N
-\frac{\I}{2m\gamma}\big)^2
+
\frac{\gamma}{ c^2}\big(N+mc^2\big)^2\vert\psi\vert^2
-
(N+mc^2){\rm I},
\label{RMCSlag}
\end{equation}
where $\beta>0$. 
Although the potential is {\it not} positive definite,
this will cause no problem when the Gauss law is
taken into account, as it  will be explained later.

The Lagrangian (\ref{RMCSlag}) is  Lorentz-invariant.
The associated equations of motion read
\begin{equation}
\begin{array}{ll}
D_\mu D^\mu\psi+\displaystyle\frac
{\partial V}{\partial\psi^*}=0,
&\hbox{non-linear Klein-Gordon eqn.}
\\[12pt]
\frac{\gamma}{ c^2}\partial_0F_{0i}+\epsilon_{ij}\partial_j F_{12}
+2\mu\epsilon_{ij}F_{0j}-J_i+{J^T}_{i}=0,\quad
&\hbox{Amp\`ere's law}
\\[12pt]
\frac{\gamma}{ c^2}\partial_i F_{0i}+2\mu F_{12}
=
\frac{\gamma}{ c^2}\big(J_0-{J^T}_{0}\big),
&\hbox{Gauss' law}
\\[12pt]
\frac{\gamma}{2c^2}\partial_\mu\partial^\mu N+
\displaystyle\frac{\partial V}{\partial N}=0
&\hbox{auxiliary eqn. for\ } $N$.
\end{array}
\label{RMCSeqmot}
\end{equation}

In a Lorentz frame where the spatial
components of the
transport current vanishes, \hfill\break
 $J^T_{\mu}=(-\frac{c^2}{\gamma}\I,0)$.
Then, using the Gauss law,  we find
for the energy
\begin{equation}
\begin{array}{ll}
H_{R}&=
\\[16pt]
&\displaystyle\int\!d^2\vx\left\{\smallover\gamma/{2c^2}\,\vec{E}\strut^2+\2B^2
+\smallover\gamma/{c^2}\,\big|D_0\psi\big|^2+\big|\vec{D}\psi\big|^2
+\smallover{\gamma^2}/{2c^4}\,\big(\partial_0N\big)^2
+\smallover\gamma/{2c^2}\,\big(\vec\nabla N\big)^2
+V\right\},
\end{array}
\end{equation}
where we used the obvious notations
$E_i=F_{0i}$, $B=F_{12}$ and we have assumed that the surface terms,
\begin{equation}
\smallover\gamma/{c^2}\,\vec\nabla\cdot\big(a_{0}\vec{E}\big)
+\mu\vec{\nabla}\times\big(a_{0}\vec{a}\big),
\end{equation}
fall off sufficiently rapidly at infinity.
To get finite energy, we require that the energy density go
to zero at infinity. Note that 
$
\vert D_{0}\psi\vert^2
$
does {\it not} go to zero at infinity, because
$$
J_{0}=(-i)\big(D_{0}\psi\psi^*-\psi(D_{0}\psi)^*\big)
$$
has to go to $J_{0}^T\neq0$ at spatial infinity.
This term
combines rather with the last two
terms in the potential.
At spatial infinity, the energy density becomes
the sum of positive terms.
Requiring that all these terms go to zero allows us to conclude that
finite energy requires 
\begin{equation}
\vert\vec{E}\vert\to0,
\qquad
B\to0,
\qquad
\vert\psi\vert^{2}\to\frac{\I}{2m\gamma},
\qquad
N\to0.
\end{equation}

Using the Bogomolny trick and the Gauss' law as written 
in Eqn. (\ref{RMCSeqmot}),
the term linear in $N$ in the potential gets absorbed.
Then the energy is re-written, for the particular value $\beta=1$, as
\begin{equation}
\begin{array}{ll}
H_{R}= \displaystyle\int&
\left\{
\smallover\gamma/{2c^2}\,\big[\vec{E}+\vec\nabla N\big]^2
+\2\big[B+\epsilon(|\psi|^2-2\vert\mu\vert N-
\frac{\I}{2m\gamma})\big]^2
\right.
\\[12pt]
&\left.
+
\smallover\gamma/{c^2}\big|D_0\psi+i(N+mc^{2})\psi\big|^2
+\big|(D_1+i\epsilon D_2)\psi\big|^2
+
\smallover\gamma^{2}/{2c^4}\big[\partial_0N\big]^2
\right\}d^2x
\\[12pt]
&
-\epsilon\big(2\vert\mu\vert mc^2-\frac{\I}{2m\gamma}\big)
\underbrace{\int B\,d^2x}_{\hbox{\small flux}},
\end{array}
\end{equation}
where $\epsilon$ is the sign of $\mu$.
The last term is topologic, labelled by the winding number,
$n$, of $\psi$. 
Due to the presence of $c^2$, it seems to be reasonable to assume that 
the coefficient in front of the magnetic flux is positive.
Then, chosing $n<0$ for $\epsilon\equiv$sign$(\mu)>0$ and
$n>0$ for $\epsilon\equiv$ sign$(\mu)<0$ respectively,
the energy admits hence the ``Bogomolny'' bound
\begin{equation}
H_R\geq
\big(2\vert\mu\vert mc^2-
\frac{\I}{2m\gamma}\big)\,2\pi \vert n\vert.
\label{RMCSbogobound}
\end{equation}

The absolute minimum  is attained by those configurations which
solve the ``Bogomolny'' equations
\begin{equation}
\begin{array}{ll}
&\partial_{0}N=0,
\\[10pt]
&\vec\nabla N+\vec{E}=0,
\\[10pt]
&
D_{0}\psi+i(N+mc^{2})\psi=0,
\\[10pt]
&
\big(D_{1}+i\epsilon D_{2}\big)\psi=0,
\\[10pt]
&
B=\epsilon\big(\frac{\I}{2m\gamma}-\vert\psi\vert^2+2\vert\mu\vert 
N\big).
\end{array}
\label{RMCSbogo}
\end{equation}
It can also be checked directly that the solutions of these 
 equations  solve the second-order field equations 
(\ref{RMCSeqmot}), when the  gauge fields are static and the matter field is of the form 
$$
\psi=e^{-imc^2t}\times\hbox{(static)}, 
$$

Eqns (\ref{RMCSbogo}) equations are similar to those of by Lee et al., and could be studied numerically as in Ref. \cite{LLM}.
Note that, just like in the case studied by Donatis and Iengo 
\cite{DoIe},
the solutions are {\it chiral} in that the winding number and the sign 
of $\mu$ are correlated.

Let us stress that for getting a non-zero electrical field, the 
presence of a non-vanishing auxiliary field $N$ is essential.
For $N=0$ we get rather a  self-dual extension of the model of
 Paul and Khare in Ref. \cite{PaKh}, whose vortex solutions
are purely magnetic.

Now we show that the non-relativistic limit 
of our relativistic model presented above is
precisely the Manton model. To see this, let us put
\begin{equation}
\psi=\frac{1}{\sqrt{2m}}e^{-imc^{2}t}\,\phi.
\label{NRlimit}
\end{equation}
The transport current is the long-distance limit of the supercurrent,
 ${J^T}_\mu=\lim_{r\to\infty}J_\mu$. But
$
\lim_{c\to\infty}J_0/c^2=-\vert\phi\vert^{2},
$
so we have
\begin{equation}
\lim_{c\to\infty}{J^T}_0/c^2=
-\lim_{r\to\infty}\vert\phi\vert^{2}=
-\lim_{c\to\infty}\frac{\I}{\gamma}\equiv-\alpha.
\end{equation}
Then the standard procedure  
yields, after dropping the term $mc^2\I$, the 
non-relativistic expression
\begin{equation}
\begin{array}{ll}
{\cal L}_{NR}=
&-\frac{1}{2}B^2+\gamma\frac{i}{2}
\big(\phi^*D_t\phi-\phi(D_t\phi)^*\big)
-\frac{1}{2m}\big|\vec{D}\phi\big|^2
\\[8pt]
&+\mu\big(Ba_t+E_2a_1-E_1a_2\big)
-\gamma a_t-\vec{a}\cdot\vec{J}^{T}
\\[8pt]
&-\Big\{\frac{\beta}{8m}\big(\alpha-|\phi|^2+
4m\vert\mu\vert N\big)^2
-\gamma\big(\alpha-|\phi|^2\big)N\Big\}.
\end{array}
\label{RelMCSLag}
\end{equation}

Note that there is no kinetic term left for the auxiliary field 
$N$. 
It can therefore be eliminated altogether
by using its equation of motion,
\begin{equation}
4\mu^2\beta N=\big(\gamma-
\frac{\vert\mu\vert\beta}{ m}\big)
\big(\alpha-\vert\phi\vert^2\big).
\label{Neq}
\end{equation}
Inserting this into the potential, this latter becomes
\begin{equation}
\big(\frac{\gamma}{4\vert\mu\vert 
m}-\frac{\gamma^2}{8\mu^2\beta}\big)
\big(\alpha-\vert\phi\vert^2\big)^{2}.
\end{equation}
For $\alpha=1$ and $m=1$ in particular, we get precisely
the Manton Lagrangian (\ref{ManLag}) with
\begin{equation}
\lambda=\frac{2\gamma}{\vert\mu\vert}-\frac{\gamma^{2}}{\mu^{2}\beta}.
\label{RelLambda}
\end{equation}

The non-relativistic limit of the equations of motion (\ref{RMCSeqmot})
is (\ref{ManNLS}-\ref{ManFCI}-\ref{ManManSD}), as it should be.

$\bullet$
In Amp\`ere's law, the first term
$(\gamma/c^2)\partial_{0}F_{0i}$ can be dropped;
setting (\ref{NRlimit}), the relativistic current  becomes the
non-relativistic expression 
$
\vec{J}=({1/2i})\big(\phi^*\vec{D}\phi-\phi(\vec{D}\phi)^*\big);
$

$\bullet$ In Gauss' law, the first term
$(\gamma/c^2)\partial_{i}F_{0i}$ can be dropped;
the time-component of the currents behave, as already noticed, as
$$
\lim_{c\to\infty}J_0/c^2=-\vert\phi\vert^{2},
\and
\lim_{c\to\infty}{J^T}_0/c^2=
-\alpha=-1.
$$

$\bullet$ In the equation for the auxiliary field $N$ the first term
$(\gamma/c^2)\partial_\mu\partial^\mu N$
can be dropped and the $c\to\infty$ limit of $\p V/\p N=0$
is (\ref{Neq});

$\bullet$ Putting (\ref{NRlimit}) into the nonlinear Klein-Gordon equation 
and using the equation of motions 
 for $N$, a lengthy but straightforward calculation yields
the non-linear Schr\"odinger equation (\ref{ManNLS}), as expected.

Note also that, for the self-dual value 
$\beta=1$ (when $\lambda$ in (\ref{RelLambda}) 
becomes (\ref{specManlambda})), 
the non-relativistic limit of the (relativistic) 
self-dual equations (\ref{RMCSbogo}))
fixes $a_0$ and $N$ as
\begin{equation}
a_{0}=N=
\big(-\frac{\epsilon}{4\mu}+\frac{\gamma}{4\mu^2}\big)
\big(1-\vert\phi\vert^2\big).
\end{equation}
which is consistent with Eq. (\ref{Manat}). The other equations reduce in 
turn to our non-relativistic self-dual equations (\ref{ManSD}).
At last, subracting 
$mc^2I$ and taking the limit $c\to\infty$, the relativistic
Bogomolny bound (\ref{RMCSbogobound})
reduces to the non-relativistic value (\ref{ManEnBogodecomp}).

%%%%%%%%%%%%%%%%%%%%%%%%%%%%%%%%%%%
\section{Spinor vortices}\label{spinorvort}
%%%%%%%%%%%%%%%%%%%%%%%%%%%%%%%%%%%

%%%%%%%%%%%%%%%%%%%%%%%%%%%%%%%%%%%%%%%%%%%%%
\subsection{Relativistic spinor vortices}\label{RSpinor}
%%%%%%%%%%%%%%%%%%%%%%%%%%%%%%%%%%%%%%%%%%%%%

In Ref. \cite{ChoRelSpin} Cho et al. obtain, 
by dimensional reduction from Minkowski 
space,
a $(2+1)$-dimensional system. After some notational changes, their
equations read
\begin{equation}\begin{array}{l}
\2\kappa\epsilon^{\alpha\beta\gamma}F_{\beta\gamma}
=
e\big(
\bar\psi_+\gamma_+^\alpha\psi_++\bar\psi_-\gamma_-^\alpha\psi_-
\big),
\\[8pt]
\big(ic\gamma^\alpha_\pm D_\alpha-m \big)\psi_\pm
=0,
\label{Choeqns}
\end{array}
\end{equation}
where the two sets of Dirac matrices are
\begin{equation}
(\gamma^\alpha_\pm)
=
(\pm (1/c)\sigma^3, i\sigma^2,-i\sigma^1),
\label{gammamat}
\end{equation}
and the $\psi_\pm$ denote the chiral components, defined as 
eigenvectors of the chirality operator
\begin{equation}
\Gamma=\left(\begin{array}{cc}
-i\sigma_3&0
\\
0&i\sigma_3
\end{array}\right).
\label{ChiralOp}
\end{equation}
Observe that, although the Dirac equations are decoupled, the
chiral components are still coupled through the Chern-Simons 
equation.
Stationary solutions, representing purely magnetic vortices, 
are readily found \cite{ChoRelSpin}.
It is particularly interesting to construct {\it static} solutions.
For $A_0=0$ and $\partial_tA_i=0$, setting
\begin{equation}
\psi_{\pm}=e^{-imt}\left(\begin{array}{c}
F_\pm\\ G_\pm
\end{array}\right),
\end{equation}
the {\it relativistic} system (\ref{Choeqns}) becomes, for $c=1$,
\begin{equation}\begin{array}{l}
\kappa\epsilon^{ij}\partial_iA_j=-e\big(|F_+|^2+|G_-|^2\big),
\\[8pt]
\big(D_1+iD_2\big)F_\pm=0,
\qquad
\big(D_1-iD_2\big)G_\pm=0.
\end{array}
\end{equation}
Now, for $F_\pm=0$ or $G_\pm=0$, these equations are
{\it identical} to those which describe the 
non-relativistic, self-dual vortices of Jackiw and Pi \cite{JaPi1,JaPi2}.

%%%%%%%%%%%%%%%%%%%%%%%%%%%%%%%%%%%%%%%%%%%%%%%%%%%
\subsection{Non-relativistic spinor vortices}\label{NRSpinor}
%%%%%%%%%%%%%%%%%%%%%%%%%%%%%%%%%%%%%%%%%%%%%%%%%%%

Non-relativistic spinor vortices
can also be constructed along the same lines \cite{DHPSpinor}.
Following L\'evy-Leblond \cite{LL}, 
a non-relativistic spin $\2$ field 
$\psi=\left(\begin{array}{c}
\Phi\\\chi
\end{array}\right)$ 
where  $\Phi$ and $\chi$ are two-component `Pauli' spinors,
is described by the $2+1$ 
dimensional equations
\begin{equation}\left\{
\begin{array}{lllll}
(\vec{\sigma}\cdot\vec{D})\,\Phi
&+&2m\,\chi&=&0,
\\[8pt]
D_t\,\Phi&+&i(\vec{\sigma}\cdot\vec{D})\,\chi&=&0.
\end{array}\right.
\label{LLeqns}
\end{equation}

These 
spinors are coupled to the Chern-Simons gauge field through 
the mass (or particle) density,
$
\varrho=|\Phi|^2,
$ 
as well as through the spatial components of the current, 
\begin{equation}
\vec{J}=i\big(\Phi^\dagger\vec{\sigma}\,\chi
-\chi^\dagger\vec{\sigma}\,\Phi\big),
\label{spinorcurrent}
\end{equation}
according to the Chern-Simons equations  (\ref{relCSEL2}).
The chirality operator is still given by Eqn. (\ref{ChiralOp})
and is still conserved. Observe that
$\Phi$ and $\chi$ in Eqn. (\ref{LLeqns}) are {\it not} the chiral components 
of $\psi$; these latter are defined by 
$
\2(1\pm i\Gamma)\psi_\pm=\pm\psi_\pm.
$

It is easy to see that Eqn. (\ref{LLeqns}) 
splits into two uncoupled systems for  
$\psi_+$ and $\psi_-$.
Each of the chiral components separately describe (in 
general different) physical phenomena in $2+1$ dimensions. 
For the ease of presentation, 
we keep, nevertheless, all four components of $\psi$.

Now the current can be written in the form:
\begin{equation}
\vec{J}=\frac{1}{2im}\Big(
\Phi^\dagger\vec{D}\Phi-(\vec{D}\Phi)^\dagger\Phi\Big)
+\vec{\nabla}\times\Big(\frac{1}{2m}\,\Phi^\dagger\sigma_3\Phi\Big).
\label{spinorcurrentbis}
\end{equation}
Using the identity
$$
(\vec{D}\cdot\vec{\sigma})^2=
\vec{D}^2+eB\sigma_3,
$$
we find that the component-spinors satisfy
\begin{equation}\left\{
\begin{array}{lll}
iD_t\Phi&=&-\displaystyle\frac{1}{2m}\Big[\vec{D}^2+eB\sigma_3\Big]\Phi,
\\[12pt]
iD_t\chi&=&-\displaystyle\frac{1}{2m}\Big[\vec{D}^2+eB\sigma_3\Big]\chi
-\frac{e}{2m}\,(\vec{\sigma}\cdot\vec{E})\,\Phi.
\label{LLeqbis}
\end{array}\right.
\end{equation}

Thus, $\Phi$ solves a \lq Pauli equation', while   
$\chi$ couples through the term,
$\vec{\sigma}\cdot\vec{E}$. 
Expressing $\vec{E}$ and $B$ through the 
Chern-Simons equations (\ref{Gauss}-\ref{FCI}) 
and inserting into our equations, 
we get finally 
\begin{equation}\left\{
\begin{array}{ll}
iD_t\Phi=
&\Big[-\displaystyle\frac{1}{2m}\,\vec{D}^2
+\displaystyle\frac{e^2}{2m\kappa}\,|\Phi|^2\,\sigma_3
\Big]\Phi,
\\[12pt]
iD_t\chi=
&\Big[-\displaystyle\frac{1}{2m}\,\vec{D}^2
+\displaystyle\frac{e^2}{2m\kappa}\,|\Phi|^2\,\sigma_3
\Big]\chi
-\displaystyle\frac{e^2}{2m\kappa}\,\big(
\vec{\sigma}\times\vec{J}\big)\Phi.
\end{array}\right.
\label{LLeqtris}
\end{equation}

If the chirality of $\psi$ is restricted to $+1$ (or $-1$),
this system describes 
non-relativistic spin $+\2$ ($-\2$) fields interacting with a 
Chern-Simons gauge field. 
Leaving the chirality of $\psi$ unspecified, it 
describes {\it two} spinor fields of spin $\pm\,\2$, 
interacting with each other
and the Chern-Simons gauge field.

Since the lower component is simply
$
\chi=-(1/2m)(\vec{\sigma}\cdot\vec{D})\Phi,
$
it is enough to solve the
$\Phi$-equation.
For 
\begin{equation}
\Phi_+=\left(\begin{array}{c}
\Psi_+\\
0\end{array}\right)
\and
\Phi_-=\left(\begin{array}{c}
0\\
\Psi_-\end{array}\right)
\end{equation}
respectively --- which amounts to working with the $\pm$ chirality
components --- the \lq Pauli' equation in (\ref{LLeqtris})
reduces to
\begin{equation}
iD_t\Psi_\pm=
\Big[-\frac{\vec{D}^2}{2m}
\pm\lambda\,(\Psi_\pm^\dagger\Psi_\pm)\Big]\Psi_\pm,
\qquad
\lambda\equiv\frac{e^2}{2m\kappa},
\label{NLSvort}
\end{equation}
which again (\ref{NLS}), but with non-linearity
$\pm\lambda$, {\it half} of the 
special value $\Lambda$ in (\ref{NRBog}), used by Jackiw and Pi.
For this reason, our solutions (presented below) 
will be {\it purely magnetic}, ($A_t\equiv0$), unlike in the case
studied by Jackiw and Pi. 

In detail, let us consider the static system
\begin{equation}\left\{\begin{array}{l}
\Big[-\frac{1}{2m}(\vec{D}^2+eB\sigma_3)-eA_t\Big]\Phi=0,
\\[12pt]
\vec{J}=-\displaystyle\frac{\kappa}{e}\vec\nabla\times A_t,
\\[12pt]
\kappa B=-e\varrho,
\end{array}\right.
\label{staticvortSD}
\end{equation}
and try the first-order Ansatz
\begin{equation}
\big(D_1\pm iD_2\big)\Phi=0
\label{vortSDAns}
\end{equation} 
that allows us to replace $\vec{D}^2=D_1^2+D_2^2$
by $\mp eB$, then the first
equation in (\ref{staticvortSD}) can be written as
\begin{equation}
\Big[-\frac{1}{2m}eB(\mp 1+\sigma_3)-eA_t\Big]\Phi=0,
\label{vortA_t}
\end{equation}
while the current is 
\begin{equation}
\vec{J}
=
\frac{1}{2m}\vec\nabla\times\Big[\Phi^\dagger(\mp1+\sigma_3)\Phi\Big].
\label{spinSDcur}
\end{equation}

Now, due to the presence of $\sigma_3$, both
Eqn. (\ref{spinSDcur}) and the second equation in (\ref{staticvortSD}) can be solved
with a {\it zero} $A_t$ and $\vec{J}$: 
by choosing 
$\Phi\equiv\Phi_+$ 
($\Phi\equiv\Phi_-$) for the upper (lower) cases respectively
makes $(\mp 1+\sigma_3)\Phi$ vanish. 
(It is readily seen from Eqn. (\ref{vortA_t}) that any solution has
a definite chirality).
The remaining task is to solve the first-order conditions
\begin{equation}
(D_1+iD_2)\Psi_+=0,\qquad{\rm or}\qquad(D_1-iD_2)\Psi_-=0,
\label{spinorSD}
\end{equation}
which is done in the same way as before~:
\begin{equation}\vec{A}=
\pm\frac{1}{2e}\vec{\nabla}\times\log\varrho+\vec{\nabla}\omega,
\qquad
\bigtriangleup\log\varrho=\pm\frac{2e^2}{\kappa}\varrho.
\label{spinorA}  
\end{equation}

A normalizable solution is obtained for
$\Psi_+$ when $\kappa<0$, and for $\Psi_-$ when $\kappa>0$.
(These correspond to attractive non-linearity
in Eqn. (\ref{NLSvort})). 
The lower components  vanish in both cases, as seen from
the $\chi$-equation
\begin{equation}
\chi=-\frac{1}{ 2m}(\vec{\sigma}\cdot\vec{D})\Phi.
%\label{chieq}
\end{equation} 
Both solutions only involve 
{\it one} of the $2+1$ dimensional spinor fields $\psi_\pm$, depending 
on the sign of $\kappa$.

The physical properties such as symmetries and conserved quantities can be studied by noting that our 
equations are in fact obtained by variation of
the $2+1$-dimensional action given in \cite{DHPSpinor},
which can also be used to 
show that the coupled L\'evy-Leblond --- Chern-Simons system is,
just like its scalar counterpart, Schr\"odinger symmetric \cite{JaPi2}.

A conserved energy-momentum tensor can be constructed and used to
derive conserved quantities \cite{DHPSpinor}. One finds that
the `particle number' $N$ determines the actual values of all 
the conserved charges: for the radially symmetric solution, e.g.,
the magnetic flux, $-eN/\kappa$, 
and the mass, ${\cal M}=mN$,  
are the same as for the scalar soliton of \cite{JaPi2}. 
The total angular 
momentum, however, can be shown to be $I=\mp N/2$,
{\it half} of the corresponding value 
for the scalar soliton. 
As a consequence of self-duality,
our solutions have {\it vanishing energy}, just like the ones of 
Ref. \cite{JaPi2}. 
 
%%%%%%%%%%%%%%%%%%%%%%%%%%%%%%%%%%%%%%%%%%%%
%\subsubsection{The non-relativistic limit}
%%%%%%%%%%%%%%%%%%%%%%%%%%%%%%%%%%%%%%%%%%%%

It is worth mentionning that our non-relativistic spinor
model here can in fact be derived from the relativistic theory
of Cho et al \cite{ChoRelSpin}.
Putting
\begin{equation}
\psi_+=e^{-imc^2t}\left(\begin{array}{c}
\Psi_+\\\widetilde{\chi}_+
\end{array}\right)
\and
\psi_-=e^{-imc^2t}\left(\begin{array}{c}
\widetilde{\chi}_-\\\Psi_-
\end{array}\right),
\end{equation}
their Eqn. (\ref{Choeqns}) become
\begin{equation}\left\{\begin{array}{l}
iD_t\Phi-c\vec{\sigma}\cdot\vec{D}\widetilde{\chi}=0,
\\[8pt]
iD_t\widetilde{\chi}+c\vec{\sigma}\cdot\vec{D}\Phi
+2mc^2\widetilde{\chi}
=0,
\end{array}\right.
\end{equation}
where
$
\Phi=\left(\begin{array}{c}
\Psi_+\\\Psi_-
\end{array}\right)
$
and
$\widetilde{\chi}=\left(\begin{array}{c}
\widetilde{\chi}_-
\\\widetilde{\chi}_+
\end{array}\right).
$
In the non-relativistic limit  
$$
mc^2\widetilde{\chi}>>iD_t\widetilde{\chi},
$$
so that this latter can be dropped from
the second equation. Redefining $\widetilde{\chi}$ as
$\chi=c\widetilde{\chi}$ yields precisely our
Eqn. (\ref{LLeqns}).
This also explains, why one gets the same (namely the Liouville)
equation both in 
the relativistic and the non-relativistic cases:
for static and purely magnetic fields, the terms containing $D_t$ 
are automatically zero.

%%%%%%%%%%%%%%%%%%%%%%%%%%%%%%%%%%%%%%%%%%%%%%%%%%%
%\subsubsection{Spinors on Bargman}
%%%%%%%%%%%%%%%%%%%%%%%%%%%%%%%%%%%%%%%%%%%%%%%%%%%

It is worth mentionning that the (2+1) dimensional
spinor model presented here can also be obtained
in the Kaluza-Klein-type framework of Section \ref{Bargmann}.
The L\'evy-Leblond equations (\ref{LLeqns}) arises, in particular,
as lightlike reduction of the \textit{massless Dirac
equation} for a 4-component Dirac spinor on
on ``Bargmann space'' $M$,
\begin{equation}
\D\psi=0.
\label{m0Dirac}
\end{equation}
This framework allows one to
rederive the Schr\"odinger symmetry of the spinor system
along the same lines as in the scalar case \cite{DHPSpinor}.

%%%%%%%%%%%%%%%%%%%%%%%%%%%%%%%%%%%%%%%%%%%%%%%%%%%
\subsection{Spinor vortices in nonrelativistic Maxwell-Chern-Simons theory}\label{SpinVort}
%%%%%%%%%%%%%%%%%%%%%%%%%%%%%%%%%%%%%%%%%%%%%%%%%%%

Now we generalize our construction to 
non-relativistic Maxwell-Chern-Simons theory of Manton's type.
Let $\Phi$ denote a 2-component Pauli spinor.
We posit the following equations of motion.
\begin{equation}\left\{
\begin{array}{ll}
i\gamma D_t\Phi=-\frac{1}{2}\big[\vec{D}^2+B\sigma_3\big]\Phi
&\hbox{Pauli eqn.}
\\[8pt]
\epsilon_{ij}\partial_{j}B
=
J_{i}-J^{T}_{i}+2\mu\,\epsilon_{ij}\,E_j\qquad\quad
&\hbox{Amp\`ere's eqn.}
\\[8pt]
2\mu B=\gamma\big(1-|\Phi|^2\big)
&\hbox{Gauss' law}
\end{array}\right.\label{spinorEqns}
\end{equation}
where the current is now 
\begin{equation}
\vec{J}=\frac{1}{2i}\Big(
\Phi^\dagger\vec{D}\Phi-(\vec{D}\Phi)^\dagger\Phi\Big)
+\vec{\nabla}\times
\Big(\frac{1}{2}\,\Phi^\dagger\sigma_3\Phi\Big).
\end{equation}
The system is plainly non-relativistic, 
and it admits self-dual vortex solutions, 
as we show now.
The transport current can again be eliminated by a
galilean boost. For fields which are static in
the frame where $\vec{J}^T=0$, the equations of motion become
\begin{equation}\left\{\begin{array}{ll}
&\big[{\2}(\vec{D}^2+B\sigma_3)+\gamma a_t\big]\Phi=0,
\\[8pt]
&\vec\nabla\times B=
\vec{J}+2\mu\,\vec\nabla\times a_t,
\\[8pt]
&2\,(\frac{\mu}{\gamma})B=1-\Phi^\dagger\Phi.
\end{array}\right.
\label{staticSpinor}
\end{equation}
Now we attempt to solve these equations
by the first-order Ansatz
\begin{equation}
\big(D_1\pm iD_2\big)\Phi=0.
\label{SDtris}
\end{equation} 
Eqn. (\ref{SDtris}) implies that
\begin{equation}
\vec{D}^2=\mp B
\and
\vec{J}
=
{\2}\vec\nabla\times\Big[\Phi^\dagger(\mp1+\sigma_3)\Phi\Big],
\end{equation}
so that the Pauli equation in (\ref{staticSpinor}) requires
\begin{equation}
\Big[(\mp 1+\sigma_3)B+2\gamma a_t\Big]\Phi=0.
\label{statPauli}
\end{equation}
Let us decompose $\Phi$ into chiral components,
\begin{equation}
\Phi=\Phi_{+}+\Phi_-
\where
\Phi_+=\left(\begin{array}{c}
0\\
\chi
\end{array}\right),
\qquad
\Phi_-=\left(\begin{array}{c}
\varphi\\0
\end{array}\right).
\end{equation}
Eqn. (\ref{statPauli}) requires that $\Phi$ have a definite chirality.
One possibility would be $\Phi_+=0$
for the upper sign, and $\Phi_-=0$ for the lower sign.
In both cases,  $a_t$ would  vanish. These choices 
are, however,
 seen to be inconsistent with Amp\`ere's law.

Curiously, there is another possibility~: one can have
\begin{equation}
a_t=\pm\smallover1/\gamma\,B,
\quad\hbox{and}\quad
\left\{\begin{array}{lll}
\Phi_-=0\quad&\hbox{i.e.}\;\Phi\equiv\Phi_+\quad
&\hbox{for the upper sign}
\\[12pt]
\Phi_+=0\quad&\hbox{i.e.}\;\Phi\equiv\Phi_-\quad
&\hbox{for the lower sign}
\end{array}\right..
\end{equation}
Then
$
\vec{J}
=
\mp\vec\nabla\times\big|\Phi_{\pm}\big|^2,
$
so that Amp\`ere's law requires
\begin{equation}
\vec{\nabla}\times\Big(\big[1\mp\smallover{2\mu}/{\gamma}\big]B
\pm\big|\Phi_{\pm}\big|^2\Big)=0.
\label{Ampbis}
\end{equation}
But now $\big|\Phi_{\pm}\big|^2=\big|\Phi\big|^2=
1-(2\mu/\gamma)B$
by the Gauss law, so that (\ref{Ampbis}) holds when
\begin{equation}
\alpha\equiv\pm\frac{\gamma}{\mu}=4.
\label{alfa}
\end{equation}
In conclusion, for the particular value
(\ref{alfa}), the second-order field equations
can be solved by solving one or the
other of the first-order equations in (\ref{staticSpinor}).
These latter conditions fix moreover the gauge potential as
\begin{equation}
\vec{a}=\pm\2\vec{\nabla}\times\log\varrho+\vec{\nabla}\omega,
\qquad
\varrho\equiv\big|\Phi\big|^2
=\big|\Phi_{\pm}\big|^2  
\end{equation}
and then the Gauss law yields
\begin{equation}
\bigtriangleup\log\varrho=4(\varrho-1),
\end{equation}
which is again the ``Liouville-type'' equation (\ref{ManLioutype}) 
studied before. Note that the sign, the same for both choices,
is automatically positive, as $\alpha=4$.

The equations of motion (\ref{spinorEqns}) can be derived from the Lagrangian
\begin{equation}
\begin{array}{ll}
{\cal L}=
&-\frac{1}{2}B^2+\displaystyle\frac{i\gamma}{2}
\big[\Phi^\dagger(D_t\Phi)-(D_t\Phi)^\dagger\Phi\big]
-\displaystyle\frac{1}{2}(\vec{D}\Phi)^\dagger(\vec{D}\Phi)
\\[12pt]
&+ \displaystyle\frac{B}{2}\Phi^\dagger\sigma_3\Phi
+\mu\big(Ba_t+E_2a_1-E_1a_2\big)
-\gamma a_t-\vec{a}\cdot\vec{J}^{T}.
\end{array}
\end{equation}
Then, in the frame where $\vec{J}^T=0$, 
the energy is
\begin{equation}
H=\frac
{1}{2}\int\left\{
B^2
+\big|\vec{D}\Phi\big|^2
-B\,\Phi^\dagger\sigma_3\Phi
\right\}\,d^2\vec{x}.
\end{equation}
Using the identity
\begin{equation}
\big|\vec{D}\Phi\big|^2=\big|(D_{1}\pm iD_{2})\Phi\big|^2
\pm B\,\Phi^\dagger\Phi %+ \hbox{surface terms}
\end{equation}
(valid up to surface terms), the energy is rewritten as
$$
H=
\2\,\int\left\{B^2
+\big|(D_{1}\pm iD_{2})\Phi\big|^2
-B\Big[\Phi^\dagger(\mp1+\sigma_3)\Phi\Big]
\right\}\,d^2\vec{x}.
$$
Eliminating $B$ using the Gauss law, we get finally,
for purely chiral fields, $\Phi=\Phi_{\pm}$, 
\begin{equation}
H=
\2\,\int\left\{
\big|(D_{1}\pm iD_{2})\Phi_{\pm}\big|^2
+\frac{\gamma}{4\mu}\big[\mp4+
\frac{\gamma}{\mu}\big]
\big(1-\vert\Phi_{\pm}\vert^2\big)^{2}
\right\}\,d^2\vec{x}
\pm\,\int B\,d^2\vec{x}.
\end{equation}
The last integral here yields the topological charge
$\pm2\pi n$.
The integral is positive definite when $\pm\gamma/\mu\geq4$,
depending on the chosen sign, yielding the Bogomolny bound
\begin{equation}
H\geq2\pi\vert n\vert.
\end{equation} 
The Pauli term  hence  {\it doubles} 
the Bogomolny bound with respect to the scalar case. 
The bound can be saturated when $\pm\gamma/\mu=4$
and the self-dual equations (\ref{SDtris}) hold.

%%%%%%%%%%%%%%%%%%%%%%%%%%%%%%%%%
\section{Conclusion and outlook}
%%%%%%%%%%%%%%%%%%%%%%%%%%%%%%%%%

In this paper, we reviewed some aspects of 
 Abelian Chern-Simons theories.  For completeness,
 we would like to list a number of related issues
 not covered by us here.

First of all, much of the properties studied here 
 can be generalized to non-Abelian 
interactions \cite{NAbCS} which have, of course, many
further interesting aspects. The Jackiw-Pi vortices, 
for example, can be generalized to $SU(N)$ gauge theory
leading to generalizations of the Liouville equation.
See, e.g., Refs. \cite{JPDT,DunneLN}. 

Experimentally, superconducting
vortices arise in fact often as lattices in a
finite domain. Within the
Jackiw-Pi model, this amounts to selecting doubly-periodic 
solutions of the Liouville equations \cite{Olesen}.

The relation to similar models which arise in
condensed matter physics could also
be developped \cite{MaPaSo}.
Other interesting aspects concern is anomalous coupling 
\cite{Torres}, as well as various self-duality properties 
\cite{Proca, SDTopMass}.

Returning to the abelian context, we should mention
the study on the dynamics of vortices \cite{SoliDyn,relVortdyn,Vortdyn}.

Let us mention, in conclusion, recent work on 
vortices in the
non-commutative, ``Moyal'' field theory \cite{NCCS} as well as
 the recent review \cite{SchapVort}.

\goodbreak

%%%%%%%%%%%%

\newpage
%%%%%%%%%%%%%%%%%%%%%%%%%%%

\end{document}